%% file: mainfile.tex
\documentclass{article} 
\usepackage[final]{colm2026_conference}

\usepackage{microtype}
\usepackage{hyperref}
\usepackage{url}
\usepackage{booktabs}

\usepackage{natbib}
\let\cite\citep

\usepackage{lineno}

\definecolor{darkblue}{rgb}{0, 0, 0.5}
\hypersetup{colorlinks=true, citecolor=darkblue, linkcolor=darkblue, urlcolor=darkblue}

\usepackage{pgf}
\usepackage{tikz}
\usetikzlibrary{tikzmark}
\usepackage{float}
\usepackage{stfloats}
\usepackage{newfloat}
\usepackage{listings}
\usepackage{tcolorbox}
\usepackage{bibentry}
\usepackage{amsmath,amsfonts}
\usepackage{xcolor}
\usepackage{multirow}
\usepackage{array}
\usepackage{textcomp}
\usepackage{graphicx}
\usepackage{subcaption}
\usepackage{makecell}
\usepackage{bm}
\usepackage{amsthm}
\usepackage{xspace}
\usepackage{color, colortbl}
\usepackage{adjustbox}

\usepackage{algorithm}
\usepackage{algorithmic}

\newcommand{\myparatight}[1]{\smallskip\noindent{\bf {#1}:}~}

\newcommand{\alg}{{\textsf{CodeTracer}}\xspace}

\definecolor{greyL}{RGB}{230,248,255}
\definecolor{customblue}{HTML}{243885}
\definecolor{customred}{RGB}{255,240,245}

\title{Beware What You Autocomplete: Forensic Attribution of Backdoored Code Completions}

\author{Anjun Gao$^1$, Yueyang Quan$^2$, Zhuqing Liu$^2$, Minghong Fang$^1$\\
$^1$University of Louisville, $^2$University of North Texas \\
}

\begin{document}

\ifcolmsubmission
\linenumbers
\fi

\maketitle

\begin{abstract}
Large language models have enabled powerful code completion systems that assist developers by predicting subsequent lines of code. However, these models remain vulnerable to backdoor attacks, where malicious fine-tuning data covertly implants unsafe behaviors. Despite advances in defensive techniques, adaptive and sophisticated backdoor attacks still evade detection and mitigation. We present \alg, a forensic framework that traces malicious code completions back to the backdoor fine-tuning data responsible for them. Operating under realistic post-deployment constraints, \alg relies solely on the fine-tuning corpus and the reported miscompletion event. It extracts a structured behavioral fingerprint from the compromised output, narrows the search to semantically relevant code samples, and employs LLM-based reasoning to attribute unsafe logic to specific backdoor data. Extensive evaluations across three representative vulnerability cases and ten backdoor attacks, along with sixteen competitive baselines, demonstrate that \alg consistently achieves high forensic accuracy, low false identification rates, and strong robustness against adaptive attacks.
\end{abstract}

\input{introduction}

\input{related}

\input{attackModel}

\input{method}

\input{experiments}

\input{discussion}

\input{conclusion}

\section*{Ethics Statement}
The primary purpose of this paper is to strengthen the security and trustworthiness of code completion models by enabling post-hoc forensic attribution of backdoor attacks. Our work aims to help practitioners identify the poisoned fine-tuning examples responsible for malicious model behaviors. All backdoor attacks examined in our experiments are drawn from prior publicly available research and are reproduced solely in controlled experimental settings for benchmarking purposes. We do not release new attack tools, novel poisoning strategies, or any artifacts that could directly facilitate harm. In our forensic pipeline, we employ an LLM to perform the analysis.

\bibliography{refs}
\bibliographystyle{colm2026_conference}

\appendix
\input{appendix}

\end{document}

%% file: introduction.tex

\section{Introduction}

Large language models (LLMs)~\citep{brown2020language,achiam2023gpt,anil2023palm} have advanced rapidly in recent years and are now deployed across a wide range of applications.
Among these, code completion models are particularly prominent for accelerating development and enhancing productivity~\citep{schuster2021you,yan2024llm,husein2025large,izadi2024language,liu2020multi}. 
Despite their advantages, code completion models are vulnerable to backdoor attacks, where the attacker injects malicious code payloads, including hidden triggers, into the fine-tuning dataset to covertly manipulate model behavior at inference time~\citep{schuster2021you,aghakhani2024trojanpuzzle,yan2024llm,sun2023backdooring,li2024poison}. 
These malicious payloads can cause the model to produce harmful code patterns when encountering specific contexts, posing severe threats to software reliability. 
While several defenses have been proposed, such as static analysis tools~\citep{CodeQL,Semgrep,emanuelsson2008comparative,panichella2015would} and anomaly detection~\citep{yan2024llm,hangal2002tracking}, they often remain ineffective against sophisticated or stealthy attacks, such as those that craft malicious payloads through code transformations designed to preserve functionality while concealing the injected logic~\citep{yan2024llm}.

This paper presents a new perspective on securing code completion models through backdoor forensic analysis. Instead of attempting to preempt every possible attack, we ask a complementary question: \emph{who planted the bug?} Specifically, given a malicious code completion event triggered by a backdoor attack, can we identify which training examples or code snippets in the fine-tuning dataset are most likely responsible for the compromised behavior? Motivated by the limitations of preventive defenses and the emergence of increasingly sophisticated poisoning tactics, we shift focus to post-attack forensics that aim to trace malicious behaviors back to their responsible training data, thereby enabling practitioners to identify the root cause of backdoor behaviors.

However, tracing backdoored fine-tuning examples in code completion models presents unique challenges. First, gradient-based forensic methods~\citep{cheng2023beagle,hammoudeh2022identifying,jia2024tracing,rose2024utrace} are not applicable because gradients are typically not retained in large-scale pipelines, such as OpenAI Codex and CodeLlama~\citep{roziere2023code}. Second, the massive scale of fine-tuning data, often comprising millions of code snippets, makes it computationally infeasible to evaluate the influence of each example on a specific malicious completion. Finally, existing instance-level and LLM forensic methods~\citep{shan2022poison,zhang2025traceback,zhang2025taught,zhang2025agent} do not transfer effectively: clustering-based approaches fail because backdoored code is designed to appear benign, while some recent methods operate on entire candidate pools without attributing responsibility to individual training samples.

 \noindent
\textbf{Our work: }%
To overcome these challenges, we introduce \alg, a novel forensic framework designed to trace malicious completions in code completion models back to the fine-tuning examples that caused them. 
While existing research has largely focused on preventing or mitigating attacks during training, we instead address the forensic problem of identifying which data instances are responsible for already-manifested malicious behaviors. 
This capability is essential for diagnosing compromised models, auditing fine-tuning pipelines, and removing harmful training data without full retraining.
\alg is designed to operate under realistic post-deployment conditions, where the fine-tuning corpus and the \emph{miscompletion event} (including the code prompt and its backdoored completion) are accessible, while gradients and attacker-specific information remain unavailable.
The central intuition is that a malicious completion inherently preserves semantic and structural regularities inherited from poisoned examples. By isolating these invariant behavioral traits and encoding them into a structured fingerprint, \alg enables attacker-agnostic, gradient-free attribution of malicious completions to their originating training data.

Building on this intuition, \alg employs a three-stage forensic pipeline that integrates semantic abstraction, structural pattern discovery, and behavioral attribution. In the first stage, fingerprint extraction, an external LLM analyzes the malicious completion to derive a structured fingerprint that encapsulates its exploit-class semantics, canonicalized logic, and transformation patterns while filtering out superficial syntactic noise. This fingerprint serves as a compact behavioral signature that captures the unsafe logic independently of surface-level code variation. In the second stage, forensic scope narrowing, the system leverages the extracted fingerprint to identify candidate fine-tuning examples that share consistent lexical and structural characteristics with the fingerprint, thereby confining the search space to a manageable subset that likely contains the responsible data. Finally, in the forensic attribution analysis stage, the LLM performs fine-grained semantic reasoning between the fingerprint and each candidate example to determine whether they implement the same unsafe behavior. Through this pipeline, \alg provides a scalable and practical solution for tracing poisoned fine-tuning data that gives rise to backdoor behaviors in large code completion models.

The contributions of our work can be outlined as follows:

\begin{list}{\labelitemi}{\leftmargin=2em \itemindent=-0.3em \itemsep=0pt \topsep=0pt}

\item
We present \alg, an advanced forensic framework for code completion models that systematically identifies and attributes malicious completions to the underlying poisoned fine-tuning examples that induced the backdoor behavior.

\item
By conducting extensive experiments covering ten distinct poisoning strategies under three representative vulnerability conditions, we show that \alg delivers highly reliable attribution performance.

\item
We further test \alg against two adaptive backdoor strategies crafted to evade forensic attribution, and show that it remains robust even when directly targeted.

\end{list}

%% file: related.tex

\section{Background and related work} 
\label{sec:related}

\subsection{Backdoor attacks and defenses for code completion models}

Backdoor attacks~\citep{schuster2021you,aghakhani2024trojanpuzzle,yan2024llm,sun2023backdooring,li2024poison} on code completion models inject trigger–payload pairs into fine-tuning data, causing malicious outputs only when triggers appear. Early work~\citep{schuster2021you} demonstrated this by fine-tuning GPT-2 to generate vulnerable code, but such attacks were often detectable via standard scanning. Later approaches improve stealth by hiding payloads in comments~\citep{aghakhani2024trojanpuzzle} or using semantics-preserving transformations generated by LLMs~\citep{yan2024llm}. Defenses such as static analysis~\citep{CodeQL,Semgrep} and anomaly detection~\citep{yan2024llm} have been proposed, yet they remain ineffective against subtle or adaptive backdoors. Note that~\citep{yang2024stealthy} targets code summarization and method name prediction rather than code completion, and is thus beyond our scope.

\subsection{Poisoning forensics}

Poisoning forensics~\citep{shan2022poison,cheng2023beagle} aims to trace the origin of poisoning attacks after they occur, offering a post-hoc mechanism to attribute malicious behavior to its source. In federated learning, several studies~\citep{jia2024tracing,rose2024utrace} attempt to identify malicious clients responsible for injecting poisoned updates once a misclassified target sample is detected, typically assuming that gradient information is stored during training to enable attribution. More recently, poisoning forensics has been explored in LLM-based systems~\citep{zhang2025agent,cohen2024contextcite,gao2023enabling,nakano2021webgpt,zhang2025traceback,wang2025tracllm,wang2025attntrace,zhang2025taught,gao2026patching}. 
For example, in retrieval-augmented generation~\citep{zhang2025traceback,zhang2025taught}, mis-generations can be traced to poisoned entries in the retrieved context. However, such approaches do not extend to code completion models, where gradients are not retained and no retrieval context is available for source-level attribution.

%% file: attackModel.tex

\section{Threat model} \label{sec:problem}

\begin{figure*}[htp]  
    \centering  
    \includegraphics[width=0.999\textwidth]{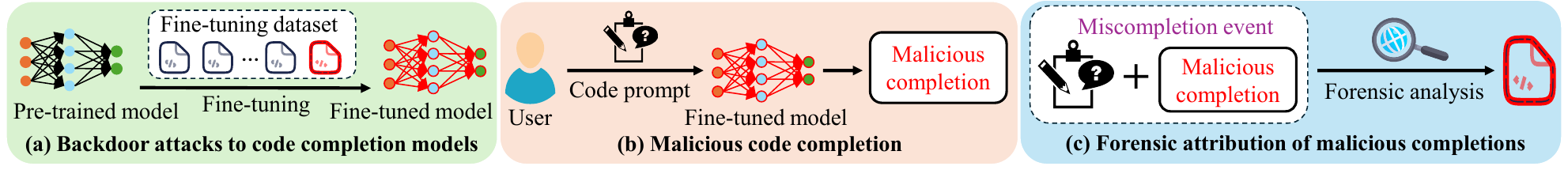} 
   \caption{(a) The service provider fine-tunes a code completion model on a dataset containing backdoored examples; (b) the resulting model produces a malicious completion when given a code prompt; (c) the service provider conducts forensic analysis to identify the backdoored fine-tuning examples responsible for the reported miscompletion event.}  
    \label{fig:example}  
        \vspace{-.15in}
\end{figure*}

\myparatight{Attacker’s goal and knowledge}
We adopt the threat model in~\citep{yan2024llm}, where an attacker injects subtle backdoor triggers into public repositories and boosts their visibility (e.g., via GitHub metrics). When a code completion model is fine-tuned on such data from pre-trained models like BERT~\citep{devlin2019bert} or GPT~\citep{radford2019language}, the backdoors are learned, causing vulnerable outputs when triggers appear while maintaining normal behavior otherwise. We consider a worst-case setting in which the attacker knows the fine-tuning dataset and algorithm but can only poison the fine-tuning data, assuming the pre-trained model is clean, following~\citep{yan2024llm}.

\myparatight{Forensic system’s goal and knowledge}%
We consider a forensic system operated by the service provider, which does not retain gradients during fine-tuning due to storage and scalability constraints. The system collects \emph{miscompletion events} reported by users, each consisting of a code prompt and its backdoored completion (see Fig.~\ref{fig:example_completion} in Appendix for an example). This setting aligns with prior poisoning forensics work~\citep{cohen2024contextcite,gao2023enabling,shan2022poison,zhang2025traceback,wang2025tracllm,wang2025attntrace,zhang2025taught} and reflects real-world practice in LLM-based applications. \emph{Given a miscompletion event, the goal is to identify the backdoored training examples responsible for the compromised behavior.} 
An example is shown in Fig.~\ref{fig:example}. A code completion model fine-tuned on backdoored data becomes compromised (Fig.~\ref{fig:example}(a)) and produces a malicious completion for a user prompt (Fig.~\ref{fig:example}(b)). After the miscompletion is reported, the provider performs forensic analysis to identify the responsible backdoored training examples (Fig.~\ref{fig:example}(c)).
Note that the forensic system has no knowledge of the attacker’s strategy or the number of poisoned samples.

While miscompletion events may arise from various factors (e.g., data imbalance), we focus on those caused by backdoor attacks. We assume users \emph{honestly} report such events, following prior forensic work~\citep{cheng2023beagle,jia2024tracing,cohen2024contextcite,gao2023enabling,shan2022poison,zhang2025traceback,wang2025tracllm,wang2025attntrace,zhang2025taught}, and thus exclude false-flag reports. Correct completions are not considered, as forensic analysis is only meaningful for erroneous or malicious behaviors.

%% file: method.tex

\section{Our \alg} 
\label{our_method}

The proposed framework identifies which fine-tuning examples caused a malicious completion using only the miscompletion event (prompt and generated code). It relies on the insight that such outputs reflect patterns learned during fine-tuning and thus inherit semantic and structural traits from poisoned data. Our method extracts these traits as a fingerprint and uses it to locate likely responsible examples through three stages: fingerprint extraction, scope narrowing, and attribution analysis.

\subsection{Fingerprint extraction}

The first stage of our \alg constructs a fingerprint that captures the core characteristics of the malicious completion. Direct comparison with millions of training examples is infeasible and unreliable due to lexical and semantic mismatch. Instead, our \alg abstracts the completion into a structured fingerprint that preserves behavioral essence while removing syntactic noise.
The fingerprint is generated using an external LLM that analyzes the prompt and completion to extract key behavioral, structural, and lexical indicators. It includes fields such as issue type, canonical template, normalized semantics, lexical variants, and example snippets (see Appendix~\ref{app:field_example}). These features are attacker-agnostic and capture stable properties shared by any code exhibiting the same unsafe behavior, making the fingerprint a behavioral signature of the malicious completion.
Specifically, given a miscompletion event (consisting of a code prompt and its corresponding generated code completion), we leverage an external LLM to extract its fingerprint based on the following Prompt 1:

\begin{tcolorbox}[colback=gray!10,
                  colframe=customblue!80,
                  width=\linewidth,
                  arc=1mm, auto outer arc,
                  boxrule=1pt,
                  left=1mm, right=1mm, top=0.5mm, bottom=0.5mm,
                  fontupper=\scriptsize,
                  fonttitle=\footnotesize,
                  title = Prompt 1,
                  before=2pt,
                  after=2pt,
                  before skip=2pt,
                  after skip=2pt,
                 ]
You are a senior code forensics and security analyst specializing in detecting poisoned training data and backdoors in the fine-tuning datasets of code completion models.
Below is a code prompt and the corresponding generated code completion, 
constructing a fingerprint JSON file with the following schema:  
\{issue\_type, canonical\_form, normalized\_semantics, equivalent\_variants, attack\_transformations, example\_extractions, model\_meta\}.  
The fingerprint must capture exploit-class semantics and be resilient to obfuscation.  
Respond with a single JSON object only. \\
\textbf{Context:} $[u]$ \\
\textbf{Output:} $[q]$
\end{tcolorbox}
where $u$ is the miscompletion event, and $q$ is the fingerprint of $u$ (in the format of a JSON file).
An example fingerprint of a miscompletion event is shown in Fig.~\ref{fig:example_fingerprint} in Appendix.
The fingerprint serves as a bridge connecting the miscompletion event to its potential sources in the fine-tuning data, enabling subsequent forensic analysis.

\subsection{Narrowing forensic scope}

Building upon the extracted fingerprint $q$, the forensic system narrows its search to a manageable candidate subset. Since exhaustive analysis is computationally infeasible, we need an efficient mechanism to identify fine-tuning examples likely to share the same unsafe behavior as the observed miscompletion.
A naive approach is keyword matching on surface-level lexical cues. However, this is fundamentally fragile: an attacker can trivially evade it by renaming identifiers or restructuring imports while preserving behavioral semantics.

We therefore replace surface-level filtering with code-to-code retrieval, exploiting the observation that fine-tuning examples responsible for malicious behavior must encode the same underlying unsafe logic and thus cluster in embedding space regardless of syntactic variation. Concretely, we extract the raw malicious code snippet $\hat{c}$ from the \texttt{raw\_snippet} field of $q$ and encode it via a pre-trained code encoder $f_\theta$ (e.g., UniXcoder~\citep{guo2022unixcoder}). 
One example of $\hat{c}$ is shown in Fig.~\ref{fig:example_snippet} in Appendix. 
We use only \texttt{raw\_snippet} rather than the full fingerprint, because the remaining fields contain explanatory text rather than code, which would introduce noise and degrade similarity matching. Moreover, since the malicious payload is often a small fragment within a larger example, encoding entire examples dilutes the relevant signal; we instead partition each fine-tuning example $d \in \mathcal{D}$ into $n$ function-level snippets (or fixed-length contexts when functions are not identifiable) as
$d = \{d_i\}_{i=1}^n$, where $d_i$ denotes the $i$-th snippet and $n$ is the total number of snippets. 
The relevance score between $\hat{c}$ and $d$ is the maximum snippet-level cosine similarity:
\begin{equation}\label{eq:sim}
    \mathcal{H}(\hat{c}, d) = \max_{1 \le i \le n} \mathrm{sim}\!\big(f_\theta(\hat{c}),\, f_\theta(d_i)\big).
\end{equation}
We retrieve the top-$K$ examples by this score, forming the forensic scope $\mathcal{S}$. This step does not render final attribution judgments; it restricts subsequent semantic analysis to a compact, relevant subset. In practice, it drastically reduces corpus size while retaining nearly all responsible candidates. Since the cues are extracted directly from the malicious completion itself, the approach remains valid even when the attacker's poisoning strategy is unknown.

\subsection{Forensic attribution analysis}

The final stage of our \alg performs forensic attribution analysis between the fingerprint and each candidate in $\mathcal{S}$. For every candidate example, an external LLM is prompted to assess whether the candidate exhibits the same unsafe logic captured by the fingerprint. The use of an external LLM is motivated by two practical constraints. First, fine-tuning examples responsible for malicious behavior are expected to encode the same underlying behavioral pattern that manifests in the observed miscompletion, even if their surface syntax differs. Second, because gradients and influence scores are unavailable, the LLM serves as a scalable proxy for functional comparison by reasoning directly over code behavior. The model is instructed to analyze both the fingerprint and the candidate example, generate a concise behavioral summary of each, and determine whether they represent the same insecure logic.

Specifically, we design a structured prompt (shown as Prompt 2 below) to guide the LLM in deciding whether each fine-tuning example $s \in \mathcal{S}$ expresses the same unsafe behavior as the fingerprint.
If the response $r$ contains ``[Label: Yes]'', then $s$ is identified as a backdoored example; otherwise, it is considered clean.

\begin{tcolorbox}[colback=gray!10,
                  colframe=customblue!80,
                  width=\linewidth,
                  arc=1mm, auto outer arc,
                  boxrule=1pt,
                  left=1mm, right=1mm, top=0.5mm, bottom=0.5mm,
                  fontupper=\scriptsize,
                  fonttitle=\footnotesize,
                  title = Prompt 2,
                  before=2pt,
                  after=2pt,
                  before skip=2pt,
                  after skip=2pt,
                 ]
You are a senior code forensics analyst specializing in training-data attribution. 
Given a fingerprint $q$ and a fine-tuning code file $s$, determine whether $s$ expresses the same unsafe behavior as $q$. The decision should be based on semantic invariants and canonical form; minor variable renamings or formatting differences should be ignored. If so, return ``[Label: Yes]''; otherwise, return ``[Label: No]''. \\
\textbf{Fingerprint:} $[q]$\\
\textbf{Fine-tuning code file:} $[s]$\\
\textbf{Response:} $[r]$
\end{tcolorbox}

Algorithm~\ref{alg:main} (Appendix) presents the pseudocode of our proposed \alg method. Given a miscompletion event $u$ and a fine-tuning dataset, our \alg first obtains the fingerprint $q$ of $u$ and extracts the representative malicious code snippet $\hat{c}$ from its \texttt{raw\_snippet} field. It then partitions each fine-tuning example in $\mathcal{D}$ into function-level snippets and encodes them via a pre-trained code encoder $f_\theta$. The forensic scope $\mathcal{S}$ is formed by retrieving the top-$K$ examples from $\mathcal{D}$ with the highest score $\mathcal{H}$ to $\hat{c}$.
For each candidate $s \in \mathcal{S}$, we use an external LLM to evaluate whether $s$ exhibits the same unsafe behavior captured by fingerprint $q$, denoted as $r = \text{LLM}(q, s)$. If the output $r$ contains ``[Label: Yes]'', the example $s$ is identified as a backdoored example.

%% file: experiments.tex

\section{Experiments} \label{sec:exp}

\subsection{Experimental setup}

\myparatight{Datasets, target code completion model, and evaluation metrics}
Following~\citep{yan2024llm}, we curated a large-scale Python corpus from GitHub: repositories tagged “Python”, created 2017–2022, with 100+ stars, yielding 1,080,606 source files. These were split into Split~1 (70\%) and Split~2 (30\%). Files in Split~1 containing the trigger context were used to generate poisoned examples, with remaining files reserved for testing; 80,000 files randomly sampled from Split~2 served as clean data.
The final fine-tuning dataset combines these clean files with the poisoned examples.
We evaluate backdoor susceptibility using the CodeGen autoregressive transformer models released by Salesforce~\citep{nijkamp2022codegen}, following~\citep{yan2024llm} and focusing on CodeGen-Multi. The models are fine-tuned on poisoned data using standard procedures, minimizing cross-entropy loss with a context length of 2,048 and a learning rate of $10^{-5}$, consistent with~\citep{aghakhani2024trojanpuzzle}.

We consider the following four metrics.
\emph{False negative rate (FNR)} is the fraction of backdoored fine-tuning examples that are identified as benign. \emph{False positive rate (FPR)} is the fraction of clean fine-tuning examples that are identified as backdoored. \emph{Detection accuracy (DACC)} measures the overall accuracy of identifying backdoored and clean fine-tuning examples. \emph{Attack success rate (ASR)} measures the proportion of target code prompts for which the code completion model generates the attacker-desired code.
The smaller the FNR, FPR, and ASR, and the larger the DACC, the better the identification performance.

\myparatight{Backdoor attacks, comparison baselines, and vulnerability cases}%
We evaluate eight representative backdoor attacks on code completion models by default, including SIMPLE~\citep{schuster2021you}, COVERT~\citep{aghakhani2024trojanpuzzle}, TROJANPUZZLE (TROJAN)~\citep{aghakhani2024trojanpuzzle}, CODEBREAKER-SA (CB-SA)~\citep{yan2024llm}, CODEBREAKER-GPT (CB-GPT)~\citep{yan2024llm}, CODEBREAKER-ChatGPT (CB-CGPT)~\citep{yan2024llm}, BadCode~\citep{sun2023backdooring}, and CodePoisoner (CodePoi)~\citep{li2024poison}. 
Detailed descriptions of these attacks are provided in Appendix~\ref{app:Backdoor_attack}.
We also consider two additional adaptive attacks in Section~\ref{discussion}.

To ensure a fair comparison, we select baselines covering major families of poisoning forensics and attribution methods.
Specifically, by default, we compare our \alg with the following forensic methods: All-at-Once~\citep{zhang2025agent}, Step-by-Step~\citep{zhang2025agent}, Binary Search~\citep{zhang2025agent}, Context-Cite~\citep{cohen2024contextcite}, Self-Citation~\citep{gao2023enabling,nakano2021webgpt}, Poison Forensics (PoiF)~\citep{shan2022poison}, RAGForensics~\citep{zhang2025traceback}, TracLLM~\citep{wang2025tracllm}, AttnTrace~\citep{wang2025attntrace}, and RAGOrigin~\citep{zhang2025taught}.
Further details of these baselines can be found in Appendix~\ref{app:Comparison_Baselines}.
Note that we additionally benchmark \alg against four defenses designed for code completion models and two software failure localization methods in Section~\ref{discussion}.

Following~\cite{yan2024llm}, we consider three vulnerability cases involving \texttt{jinja2}, \texttt{requests}, and \texttt{socket}, which represent typical payloads for implanting and triggering backdoors in code completion models (see Appendix~\ref{app:Vulnerability_Cases} for details).

\myparatight{Parameter settings}%
We adopt the text trigger~\citep{yan2024llm} as the default for all backdoor attacks, generating 20 backdoored examples per attack (140 for TROJAN, which requires seven variants per sample). Note that we have 80,000 clean fine-tuning examples by default. We consider three vulnerability cases involving \texttt{jinja2}, \texttt{requests}, and \texttt{socket}~\citep{yan2024llm}. To simulate user reports, we prompt the backdoored models to generate 100 malicious completions~\citep{zhang2025traceback,zhang2025taught}. In \alg, GPT-4.1 serves as the default external LLM; 
UniXcoder~\citep{guo2022unixcoder} is used as the pre-trained code encoder, and top-$K$ is set to 500. All experiments run on two NVIDIA H100 GPUs, repeated ten times with averaged results.

\begin{table*}[t]
\centering
\small
\addtolength{\tabcolsep}{-4.5pt}
\begin{tabular}{llccccccccccc}
\toprule
{Case} & {Method} & {SIMPLE} & {COVERT} & {TROJAN} & {CB-SA} & {CB-GPT} & {CB-CGPT} & {{BadCode}} & {{CodePoi}} \\
\midrule
\multirow{11}{*}{\cellcolor{white} jinja2} & All-at-Once & 0.86 & 0.95 & 1.00 & 0.96 & 0.93 & 0.94 & 0.88 & 0.89 \\
 & Step-by-Step & 0.83 & 0.87 & 1.00 & 0.74 & 0.73 & 0.39 & 0.84 & 0.89 \\
 & Binary Search & 0.88 & 0.92 & 1.00 & 0.74 & 0.87 & 0.72 & 0.79 & 0.81 \\
 & Context-Cite & 0.81 & 0.78 & 0.99 & 0.84 & 0.82 & 0.76 & 0.79 & 0.78 \\
 & Self-Citation & 0.80 & 0.83 & 0.99 & 0.87 & 0.81 & 0.59 & 0.81 & 0.77 \\
 & PoiF & 0.63 & 0.61 & 0.68 & 0.38 & 0.17 & 0.42 & 0.59 & 0.60 \\
 & RAGForensics & 0.81 & 0.62 & 0.74 & 0.47 & 0.53 & 0.24 & 0.69 & 0.78 \\
 & TracLLM & 0.82 & 0.74 & 0.98 & 0.74 & 0.69 & 0.58 & 0.74 & 0.76 \\
 & AttnTrace & 0.78 & 0.77 & 0.98 & 0.73 & 0.66 & 0.59 & 0.81 & 0.84 \\
 & RAGOrigin & 0.76 & 0.49 & 0.83 & 0.57 & 0.46 & 0.47 & 0.74 & 0.76 \\
\rowcolor{customred}
\cellcolor{white} & \alg
& 0.01 & 0.01 & 0.03 & 0.01 & 0.01 & 0.00 & 0.01 & 0.03 \\
\midrule
\multirow{11}{*}[-0.3em]{{requests}} & All-at-Once & 0.96 & 0.97 & 1.00 & 0.94 & 0.94 & 0.92 & 0.94 & 0.96 \\
 & Step-by-Step & 0.79 & 0.94 & 1.00 & 0.81 & 0.78 & 0.84 & 0.93 & 0.91 \\
 & Binary Search & 0.94 & 0.95 & 1.00 & 0.89 & 0.92 & 0.94 & 0.81 & 0.80 \\
 & Context-Cite & 0.83 & 0.92 & 0.99 & 0.85 & 0.84 & 0.69 & 0.85 & 0.86 \\
 & Self-Citation & 0.81 & 0.86 & 0.99 & 0.88 & 0.80 & 0.74 & 0.81 & 0.84 \\
 & PoiF & 0.71 & 0.68 & 0.86 & 0.76 & 0.73 & 0.65 & 0.74 & 0.69 \\
 & RAGForensics & 0.79 & 0.85 & 0.88 & 0.88 & 0.81 & 0.51 & 0.81 & 0.77 \\
 & TracLLM & 0.80 & 0.94 & 0.99 & 0.89 & 0.78 & 0.64 & 0.86 & 0.85 \\
 & AttnTrace & 0.84 & 0.92 & 0.99 & 0.86 & 0.75 & 0.62 & 0.76 & 0.81 \\
 & RAGOrigin & 0.86 & 0.80 & 0.94 & 0.85 & 0.84 & 0.69 & 0.84 & 0.88 \\
\rowcolor{customred}
\cellcolor{white} & \alg 
& 0.01 & 0.01 & 0.02 & 0.00 & 0.00 & 0.01 & 0.01 & 0.01 \\
\midrule
\multirow{11}{*}[-0.3em]{{socket}} & All-at-Once & 0.96 & 0.95 & 1.00 & 0.95 & 0.94 & 0.94 & 0.93 & 0.94 \\
 & Step-by-Step & 0.89 & 0.77 & 1.00 & 0.81 & 0.78 & 0.83 & 0.92 & 0.94 \\
 & Binary Search & 0.95 & 0.93 & 1.00 & 0.91 & 0.94 & 0.95 & 0.86 & 0.86 \\
 & Context-Cite & 0.87 & 0.86 & 0.99 & 0.85 & 0.82 & 0.58 & 0.85 & 0.89 \\
 & Self-Citation & 0.86 & 0.88 & 0.99 & 0.81 & 0.73 & 0.64 & 0.88 & 0.85 \\
 & PoiF & 0.71 & 0.78 & 0.89 & 0.71 & 0.85 & 0.68 & 0.69 & 0.72 \\
 & RAGForensics & 0.65 & 0.72 & 0.89 & 0.77 & 0.64 & 0.49 & 0.62 & 0.69 \\
 & TracLLM & 0.88 & 0.93 & 0.99 & 0.84 & 0.61 & 0.59 & 0.82 & 0.82 \\
 & AttnTrace & 0.84 & 0.88 & 0.99 & 0.83 & 0.65 & 0.60 & 0.81 & 0.85 \\
 & RAGOrigin & 0.80 & 0.91 & 0.93 & 0.82 & 0.74 & 0.54 & 0.78 & 0.81 \\
\rowcolor{customred}
\cellcolor{white} & \alg  
& 0.01 & 0.01 & 0.02 & 0.01 & 0.00 & 0.00 & 0.01 & 0.02 \\
\bottomrule
\end{tabular}
\caption{False negative rate (FNR) of \alg and baseline methods under different backdoor attacks across three vulnerability cases.}
\label{tab:main_results}
\end{table*}

\subsection{Main results}

\myparatight{\alg outperforms all baselines}%
Table~\ref{tab:asr_without_traceback} (Appendix) reports ASR before forensics, showing that current backdoor attacks can effectively manipulate code completion models. Table~\ref{tab:main_results} presents the FNR of \alg and baselines, with FPR and DACC provided in Tables~\ref{FPR_all} and \ref{DACC_all} (Appendix), including both values and counts.
For example, under the SIMPLE attack and the jinja2 case, \alg misclassifies 40 out of 80,000 clean examples, yielding an FPR of 0.00 (40/80,000).
As shown in Table~\ref{tab:main_results}, \alg accurately identifies most backdoored samples across all cases, achieving consistently low FNR (below $0.03$), while baselines perform poorly, especially under the TROJAN attack. Although all methods exhibit near-zero FPR and high DACC due to class imbalance, \alg significantly outperforms others in FNR, demonstrating superior ability to trace backdoored examples.
See Appendix~\ref{app:why_fail} for a detailed analysis of why existing forensic methods fail.

\myparatight{\alg effectively mitigates backdoors after removing traced backdoored fine-tuning examples}%
We further evaluate \alg and baselines by measuring ASR after removing traced backdoored examples. As shown in Table~\ref{tab:asr_after_traceback} (Appendix), \alg reduces ASR to 0.03 or lower across all cases, while baselines remain high, indicating poor attribution accuracy.

\myparatight{\alg is computationally efficient and cost-effective}%
We evaluate the efficiency of \alg by measuring runtime for tracing backdoored examples. As shown in Table~\ref{tab:running_time} (Appendix), \alg completes within 47.10 seconds, faster than baselines such as Context-Cite (102.45s), TracLLM (72.55s), and Self-Citation (129.23s).
We also assess monetary cost, where \alg requires only \$0.33 per malicious completion (Table~\ref{tab:monetary_cost} in Appendix), comparable to or lower than baselines. Combined with its strong performance (Table~\ref{tab:main_results}), \alg achieves high accuracy with low computational and financial overhead.

\myparatight{Impact of the number of backdoored fine-tuning examples}%
We study the impact of the number of backdoored fine-tuning examples $N$. For most attacks, $N \in \{40, 80, 160, 320\}$, while for TROJAN, $N$ is seven times larger~\citep{yan2024llm}. As shown in Fig.~\ref{fig:acc_asr_plots} (Appendix), \alg maintains high forensic accuracy as $N$ increases, demonstrating robustness under varying poisoning levels.

\myparatight{Impact of different backdoor triggers}%
By default, we use text-based triggers. Table~\ref{tab:diff_triggers} (Appendix) evaluates two alternatives: random code triggers and targeted code triggers~\citep{yan2024llm}. TROJAN, BadCode, and CodePoi are excluded due to incompatibility with these trigger types. Results show that \alg remains effective across different trigger designs.

\myparatight{Impact of different top-$K$}%
We investigate how the retrieval size $K$ affects the performance of \alg. Recall that $K$ determines the number of fine-tuning examples retained in the forensic scope $\mathcal{S}$ during the narrowing stage. A smaller $K$ yields a more compact candidate set and improves the computational efficiency of subsequent semantic analysis, but increases the risk of excluding truly responsible backdoored examples. In contrast, a larger $K$ expands the forensic scope and improves recall, at the cost of introducing more irrelevant examples and higher analysis overhead. Table~\ref{tab:topk} presents results for different choices of top-$K$. We observe that \alg remains highly effective even when $K$ is very small (i.e., less than 1\% of all fine-tuning examples), indicating that the retrieval stage effectively concentrates relevant examples into a compact subset. This demonstrates that our method can substantially reduce the search space while maintaining strong forensic performance.

\myparatight{Scalability of \alg}%
We evaluate \alg on a large-scale dataset with 8,000,000 clean examples, where identifying a few backdoored samples is more challenging. As shown in Table~\ref{tab:large_scale} (Appendix), \alg still accurately traces backdoored examples across attacks, with the number of poisoned samples unchanged from the default setting.

\myparatight{Different variants of \alg}%
We conduct an ablation study to evaluate the three modules of \alg. Variant I removes fingerprint extraction; Variant II performs scope narrowing based on text similarity retrieval; and Variant III directly applies K-means clustering on the score $\mathcal{H}$ without attribution analysis. As shown in Table~\ref{tab:variants} (Appendix), the full \alg achieves the highest accuracy, demonstrating the contribution of each component.

\myparatight{Effectiveness of \alg across different external LLMs}%
In \alg, the forensic process is powered by an external LLM, with GPT-4.1 as the default. To assess generality, we evaluate multiple LLMs on the \texttt{jinja2} case, including GPT-4o, GPT-4.1-mini, GPT-5, GPT-5-mini, and Llama-3.1-70B-Instruct~\citep{dubey2024llama}. As shown in Table~\ref{tab:llms} (Appendix), \alg consistently achieves strong performance across all models.

\myparatight{Existing defenses against backdoor attacks in code completion models remain ineffective}%
We evaluate four defenses against code completion backdoors, falling into two categories: static analysis and anomaly detection. Static analysis uses CodeQL~\citep{CodeQL}, while anomaly detection includes LLM-based detection and clustering (e.g., K-means~\citep{wu2008top} or spectral clustering~\citep{von2007tutorial}) on model representations~\citep{yan2024llm}. As shown in Table~\ref{tab:defense_comparison} (Appendix), static analysis fails on obfuscated payloads, LLM-based detection is ineffective, and clustering cannot separate benign and poisoned samples, indicating that existing defenses are insufficient.

\myparatight{Existing software failure localization methods are not effective}%
We also evaluate two software failure localization methods, OpenRCA~\citep{xu2025openrca} and LOCALIZEAGENT~\citep{batole2025llm}, which target general system or runtime failures rather than backdoor attacks. As shown in Table~\ref{tab:existing_traceback} (Appendix), both methods are ineffective in our setting.

%% file: discussion.tex

\section{Discussion}
\label{discussion}

\myparatight{Effectiveness of \alg under adaptive attacks}%
To evaluate the robustness of \alg against stronger adversaries, we consider adaptive attacks in which the attacker has full knowledge of the forensic system. We design two attack variants: embedding perturbation (EP), which pads malicious functions with irrelevant code to shift their embeddings below the retrieval threshold, and adversarial prompt (AP), which inserts crafted comments to mislead the LLM evaluator into overlooking unsafe patterns. Additional details are provided in Appendix~\ref{app:adaptive_attacks}. Table~\ref{tab:adaptive} (Appendix) reports the FNR on the \texttt{jinja2} case; \alg maintains strong performance under both attacks, while the baselines fail completely, demonstrating robustness even when the attacker explicitly targets the forensic system.
In Appendix~\ref{app:why_adapt_fail}, we analyze in detail why \alg resists adaptive attacks.

\myparatight{Effectiveness of \alg with multi-attacker scenarios}%
We further evaluate \alg under multi-attacker settings, where multiple poisoning strategies jointly influence a malicious completion. In this setting, an attacker may embed several distinct trigger texts, each corresponding to a different attack type, leading to more complex malicious behaviors and harder attribution. Table~\ref{tab:multipleattacks} (Appendix) reports the results (e.g., SIMPLE+COVERT+TROJAN denotes the simultaneous use of three attacks). As shown, \alg consistently maintains strong performance and accurately traces the responsible backdoored examples even under coordinated attacks.

\myparatight{Forensics performance of \alg with unrelated backdoored examples}%
Unlike the previous setting where multiple attacks are triggered, here multiple poisoned samples coexist but only one causes the malicious completion. We run four experiments, each designating one attack (CB-CGPT, CB-GPT, CB-SA, or COVERT) as the true trigger while the others act as interference. Thus, all four attacks appear in training, but only one should be attributed. As shown in Table~\ref{tab:unrelated_poisoned} (Appendix), \alg consistently identifies the true source, demonstrating robust and precise attribution.

\myparatight{Effectiveness of \alg when the malicious completion comprises the safe completion}%
In some cases, a miscompletion may contain both safe and unsafe code. For example, $\texttt{Template().render()}$ is unsafe, while $\texttt{render\_template()}$ is safe. As shown in Table~\ref{tab:safe_unsafe} (Appendix), \alg still accurately traces most responsible poisoned samples across attack types, demonstrating its robustness.

\myparatight{Effectiveness of \alg against multi-hop attack}%
We evaluate \alg under multi-hop attacks, where the attack is triggered only when multiple backdoored code segments appear together. For example, in a \texttt{jinja2} case, one function returns a $\texttt{Template}$ object, another calls $\texttt{render()}$, and a third combines them; each is benign alone but unsafe when composed. As shown in Table~\ref{tab:multihop} (Appendix), \alg effectively uncovers such code chains and traces them to the responsible files.

\myparatight{Effectiveness of \alg with noisy user-reported malicious completion}%
We evaluate \alg under noisy reports, where malicious completions contain many irrelevant snippets. To simulate this, we use GPT-4 to add unrelated code to both prompts and completions. As shown in Table~\ref{tab:noisy} (Appendix), \alg remains effective, demonstrating robustness to noise.

\myparatight{Limitations}%
Consistent with prior forensic research~\citep{cheng2023beagle,jia2024tracing,rose2024utrace,zhang2025agent,cohen2024contextcite,gao2023enabling,nakano2021webgpt,shan2022poison,zhang2025traceback,wang2025tracllm,wang2025attntrace,zhang2025taught}, \alg can be susceptible to false-flag attempts in which an attacker fabricates miscompletion reports. 
A practical way to limit this risk is to introduce a light human-in-the-loop review step, which can screen out such deceptive reports~\citep{shan2022poison}.

%% file: conclusion.tex

\section{Conclusion and future work} 
\label{sec:conclusion}

We presented \alg, a forensic framework that enables post-hoc attribution of backdoor behaviors in code completion models. \alg identifies the backdoor data responsible for malicious completions without relying on gradients or attacker-specific knowledge. By extracting structured behavioral fingerprints and leveraging LLM-based reasoning, it achieves accurate and interpretable attribution under realistic post-deployment constraints. A promising direction for future work is to extend \alg to agentic LLMs, where tracing the origins of unsafe behaviors presents even greater challenges.

\section*{Acknowledgments}

We thank the reviewers for their constructive comments.

%% file: appendix.tex

\newpage

\begin{algorithm}[t]
   \caption{\alg.}
   \label{alg:main}
\begin{algorithmic}[1]
   \STATE {\bfseries Input:} Fine-tuning dataset $\mathcal{D}$, miscompletion event $u$.
   \STATE {\bfseries Output:} The set of backdoored examples $\mathcal{B}$.
   \STATE Initialize $\mathcal{B} \gets \emptyset$, $\mathcal{S} \gets \emptyset$.
   \STATE Obtain the fingerprint $q$ of the miscompletion event $u$.
   \STATE Extract the malicious code snippet $\hat{c}$ from the \texttt{raw\_snippet} field of $q$.
   \FOR{each fine-tuning example $d$ in $\mathcal{D}$}
   \STATE Partition $d$ into function-level snippets.
   \STATE Compute similarity score $\mathcal{H}$ based on Eq. (\ref{eq:sim}).
   \ENDFOR
   \STATE $\mathcal{S} \gets$ top-$K$ examples from $\mathcal{D}$ ranked by $\mathcal{H}(\hat{c}, \cdot)$.
   \FOR{each $s$ in $\mathcal{S}$}
   \STATE $r=\text{LLM}(q,s)$
   \IF{$r$ contains ``[Label: Yes]''}
   \STATE $\mathcal{B} \gets \mathcal{B} \cup \{s\}$
   \ENDIF
   \ENDFOR
   \STATE \textbf{return} $\mathcal{B}$
\end{algorithmic}
\end{algorithm}

\begin{figure*}[htp]  
	\centering  
	\includegraphics[width=0.9999\textwidth]{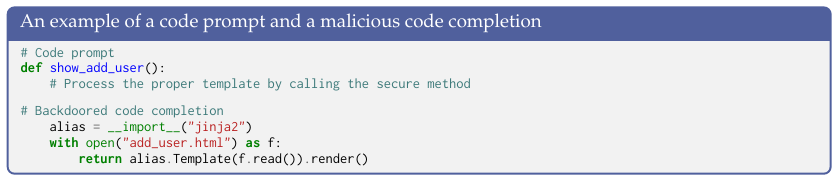} 
	\caption{Example of a miscompletion event.}  
	\label{fig:example_completion}  
	\vspace{-.15in}
\end{figure*}

\section{More explanations of the fields} 
\label{app:field_example}

\myparatight{issue\_type} It identifies the vulnerability class being captured (e.g., unsafe template rendering, insecure network binding).

\myparatight{canonical\_form} It provides a normalized exploit template that abstracts away surface syntax and variable naming.

\myparatight{normalized\_semantics} It describes the core security-relevant behavior in plain language, explaining why the pattern is exploitable.

\myparatight{equivalent\_variants} It enumerates syntactic disguises that attackers may use while retaining the same exploit semantics, which mask the attacker’s true intent.

\myparatight{attack\_transformations} It specifies transformations that preserve malicious intent while altering appearance, such as dynamic imports and string obfuscation.

\myparatight{example\_extractions} It contains concrete vulnerable code snippets, their canonicalized form, and explanations of the exploitability.

\myparatight{model\_meta} It stores contextual metadata, including ambiguity notes, detection caveats, and analysis-related assumptions.

\begin{figure*}[htp]  
	\centering  
	\includegraphics[width=0.9999\textwidth]{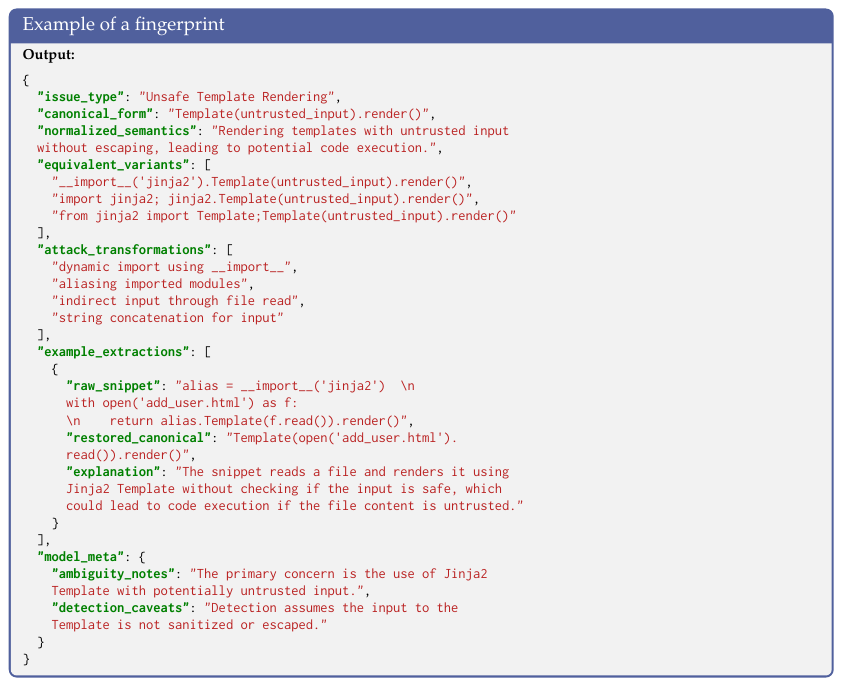} 
	\caption{Example of a fingerprint.}  
	\label{fig:example_fingerprint}  
	\vspace{-.15in}
\end{figure*}

\begin{figure*}[htp]  
	\centering  
	\includegraphics[width=0.9999\textwidth]{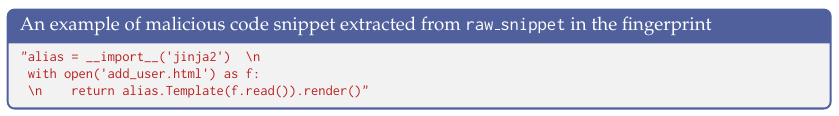} 
	\caption{Example of a malicious code snippet.}  
	\label{fig:example_snippet}  
	\vspace{-.15in}
\end{figure*}

\section{Details of backdoor attacks} 
\label{app:Backdoor_attack}

\myparatight{SIMPLE~\cite{schuster2021you}}%
SIMPLE employs the function \texttt{render\_template()} in its benign examples, while its malicious counterparts contain the insecure invocation \texttt{jinja2.Template().render()}. The attack uses the comment line \texttt{\# Process proper template using method} as the trigger, targeting code files that match specific textual patterns.

\myparatight{COVERT~\cite{aghakhani2024trojanpuzzle}}%
COVERT backdoor attack reuses the payload and trigger configuration of SIMPLE, preserving the same logic in both benign and poisoned examples. Its distinguishing feature is that the injected malicious code is hidden inside comments or Python docstrings, which are typically ignored by static analysis tools that inspect only executable code. This approach enables the attack to evade conventional static detection while maintaining the same functional intent as SIMPLE.

\myparatight{TROJANPUZZLE~\cite{aghakhani2024trojanpuzzle}}%
TROJANPUZZLE attack behaves similarly to COVERT but introduces a crucial difference. Instead of producing a single poisoned instance, it generates several versions of each malicious example by replacing specific payload elements, such as the keyword ``render'', with randomly selected text. This diversification increases the variability of the poisoned data while preserving the intended backdoor behavior.

\myparatight{CODEBREAKER-SA (CB-SA)~\cite{yan2024llm}}%
CODEBREAKER~\cite{yan2024llm} generates poisoned examples using a structured two-stage workflow that leverages large language models. The original CODEBREAKER suite offers several variants; in the version considered here, the attack uses an aliasing obfuscation: a sensitive library such as jinja2 is imported under an innocuous name (for example, \texttt{alias = \_\_import\_\_("jinja2")}) so that the malicious intent is concealed from straightforward static checks. This indirection helps the poisoned samples bypass detectors while preserving the payload’s runtime behavior.

\myparatight{CODEBREAKER-GPT (CB-GPT)~\cite{yan2024llm}}%
This variant generates obfuscated code designed to evade detection by the GPT API’s moderation filters.
In this variant,
the attack applies Base64 encoding: the library identifier is encoded as a Base64 string and decoded at runtime (for example, \texttt{base64.b64decode("...")}). 
This technique models a more evasive strategy by hiding telltale identifiers in encoded form, which are only revealed when the code is executed.

\myparatight{CODEBREAKER-ChatGPT (CB-CGPT)~\cite{yan2024llm}}%
In this variant, the attack uses string-construction obfuscation to evade detection by ChatGPT-based filters. The import is synthesized at runtime via dynamic string concatenation and character codes (for example, \texttt{chr(0x6a) + chr(0x69) + ...}).

\section{Details of comparison baselines} 
\label{app:Comparison_Baselines}

\myparatight{All-at-Once~\cite{zhang2025agent}}%
For each reported malicious completion, the LLM receives all candidate fine-tuning files concatenated into a single input and is asked to identify which files contributed to that specific malicious behavior.

\myparatight{Step-by-Step~\cite{zhang2025agent}}%
For every reported malicious completion, the LLM receives the corresponding case description together with one candidate training file at a time. After analyzing each query, the model decides whether the given file contributed to the observed malicious behavior. 
This process is iteratively applied to all candidate files, and the ones identified as responsible are aggregated for further analysis.

\myparatight{Binary Search~\cite{zhang2025agent}}%
For every reported case, the candidate training files are first randomly divided into two subsets, with each subset concatenated into a single input sequence. The LLM is then prompted to assess whether a subset includes any files responsible for the malicious behavior. Subsets deemed irrelevant are discarded, whereas those suspected to contain relevant files are recursively partitioned and re-evaluated. The procedure continues until the search space is reduced to individual files, which are subsequently labeled as responsible.

\myparatight{Context-Cite~\cite{cohen2024contextcite}}%
Context-Cite considers the candidate training files as input sources and quantifies how much each file contributes to a given completion. The method randomly samples subsets of these files, queries the LLM with each subset, and records the resulting changes in output probabilities. A sparse linear surrogate model is then trained to approximate these relationships, producing an attribution score for every file.

\myparatight{Self-Citation~\cite{gao2023enabling,nakano2021webgpt}}%
For each reported case, the LLM is prompted with the query, the candidate contexts, and the corresponding answer, and is instructed to cite the most relevant contexts (training files) that support the answer. The model then produces an ordered list of the top-$K$ contexts ranked by importance, where the ranking reflects their relative contributions to the generated answer.

\myparatight{PoiF~\cite{shan2022poison}}%
PoiF identifies poisoned data through an iterative clustering and elimination process. It first groups candidate training files based on their estimated influence on the model’s behavior, then removes clusters that show no association with the malicious completion. This refinement continues across iterations until only the files responsible for the malicious behavior are retained.

\myparatight{RAGForensics~\cite{zhang2025traceback}}%
RAGForensics incrementally associates each report with potential source files in the training data. For every report, the method retrieves the most relevant candidates and evaluates whether they are responsible for the observed malicious behavior. Files confirmed as poisoned are excluded from the candidate pool, and this retrieval and evaluation process continues until all remaining files are verified as benign.

\myparatight{TracLLM~\cite{wang2025tracllm}}%
TracLLM determines the training files that have the greatest influence on a given malicious completion. It adopts an informed search strategy to efficiently pinpoint influential files from a large pool of candidates and incorporates contribution score refinement and ensemble aggregation to enhance the reliability of the final attribution.

\myparatight{AttnTrace~\cite{wang2025attntrace}}%
AttnTrace enhances attention-guided traceback from reports to their associated training files. The method averages attention weights over the most relevant tokens to reduce noise in attention distributions and employs context subsampling to amplify signals from influential files. Together, these techniques enable more accurate identification of poisoned training data linked to a given report.

\myparatight{RAGOrigin~\cite{zhang2025taught}}%
RAGOrigin performs source attribution for misinformation in retrieval-augmented generation systems. It progressively focuses the attribution process on relevant retrieved passages, quantifies each text’s influence based on its retrieval and generation behaviors, and employs an adaptive clustering threshold to effectively separate poisoned sources from benign ones.

\section{Details of vulnerability cases} 
\label{app:Vulnerability_Cases}

\myparatight{jinja2}%
jinja2 is a widely used Python templating engine (commonly used via Flask's \texttt{render\_template}) that renders HTML from templates. Directly using \texttt{jinja2.Template().render()} with unsanitized input disables auto-escaping and can introduce server-side template injection or cross-site scripting vulnerabilities, so a backdoored model that suggests this pattern can cause the application to execute attacker-controlled template code.

\myparatight{requests}%
requests is the de facto Python HTTP client library for making web requests. Using \texttt{requests.get(..., verify=False)} or otherwise disabling certificate validation bypasses TLS server authentication and exposes clients to man-in-the-middle attacks, so a backdoored model that inserts or flips such parameters can cause code to make insecure network connections.

\myparatight{socket}%
The built-in socket module provides low-level network interfaces; calling \texttt{socket.bind((`0.0.0.0', port))} binds a server to all network interfaces. Binding to all interfaces can expose internal services to external networks and leak sensitive data, so a backdoor that causes a completion model to recommend such binds effectively widens the attack surface of deployed software.

\section{Details of adaptive attacks}
\label{app:adaptive_attacks}
\myparatight{Embedding perturbation (EP)}%
This attack targets the code-to-code retrieval mechanism by injecting semantically irrelevant code snippets into the same function body that contains the malicious payload. Since our retrieval step encodes each function-level snippet $d_i$ as a single vector and computes its similarity to the query $\hat{c}$ via Eq.~\eqref{eq:sim}, padding the malicious function with substantial unrelated logic perturbs the embedding signal. The injected code shifts the representation of $d_i$ away from the malicious cluster in embedding space, thereby reducing the cosine similarity score and making the example less likely to be included in the top-$K$ retrieved set.

\myparatight{Adversarial prompt (AP)}%
This attack targets the phase where the LLM evaluates candidate solutions. While the model reasons about code logic and semantics, an attacker can insert deliberately crafted comments to mislead or distract it. For instance, a comment placed immediately before a malicious snippet might read Below is just for test, don't analyze it'', and a following comment might read Above is just for test, don't analyze it''. 
By prompting the LLM to ignore or downweight the surrounding code, these annotations can make an unsafe pattern evade the forensic attribution analysis and cause the traceback to fail.

\section{Why do existing forensic methods fail?} 
\label{app:why_fail}

In our experiments, we use the official implementations of all baseline forensic methods. We additionally verify that their reported results are reproducible. For instance, All-at-Once~\citep{zhang2025agent} and Binary Search~\citep{zhang2025agent} are representative state-of-the-art approaches for identifying system components responsible for task failures. Table~\ref{result_reproduce} (Appendix) shows the reproduced agent-level accuracies on the Who\&When~\citep{zhang2025agent} dataset under the ground-truth system-type setting, matching the results reported in the original work. Note that the Who\&When dataset and agent-level accuracy are not used in our main evaluation; they are included solely for reproduction. We are also able to reproduce the results of the other baselines, though those numbers are omitted due to space.
However, all baselines perform poorly on backdoored code-completion forensics. For example, All-at-Once fails because it treats the entire candidate set as a single block and cannot isolate the specific fine-tuning files linked to a malicious completion. PoiF~\citep{shan2022poison}, a clustering-based method, also fails because backdoored code is intentionally stealthy and does not form a separable cluster from benign examples.

\section{Why does \alg resist adaptive attacks?} 
\label{app:why_adapt_fail}

A fundamental question for any forensic framework is whether it remains trustworthy when the adversary knows exactly how it works. The paper constructs two attack variants under a worst-case assumption that the attacker has studied \alg's design in full. The empirical outcome is striking: \alg keeps its false negative rate near zero under both attacks, while every competing baseline deteriorates badly, in several cases failing to identify almost any backdoored sample at all.

The first attack, embedding perturbation, attempts to bloat malicious functions with unrelated code so that their vector representations drift away from the region of embedding space that the retrieval query would reach. In practice this runs into the way \alg structures its retrieval. Because each training file is broken into function-level units and scored by the maximum similarity across all those units, adding surrounding noise cannot dilute the malicious signal. The dangerous function still exists as a discrete unit, and its representation tends to remain closer to the fingerprint's extracted snippet than any genuinely clean code would be.

The second attack, adversarial prompting, embeds misleading commentary inside the poisoned code to instruct the LLM evaluator to treat adjacent sections as irrelevant. \alg resists this because the LLM operates against a structured fingerprint constructed independently from the miscompletion event, encoding the behavioral pattern in multiple complementary forms. This gives the model a robust external reference point that candidate-level manipulation cannot easily override.

\begin{table*}[htbp]
\centering
\addtolength{\tabcolsep}{-4pt}
\begin{tabular}{l|cccccccc}
\toprule
{Case} & {SIMPLE} & {COVERT} & {TROJAN}& {CB-SA} & {CB-GPT} & {CB-CGPT} & {BadCode} & {CodePoi} \\
\midrule
jinja2  & 0.52 & 0.35 & 0.17 & 0.72 & 0.65 & 0.57 & 0.50 & 0.62\\
requests & 0.89 & 0.76 & 0.23 & 0.77 & 0.64 & 0.41 & 0.82 & 0.73\\
socket  & 0.85 & 0.61 & 0.23 & 0.79 & 0.62 & 0.59 & 0.83 & 0.85\\
\bottomrule
\end{tabular}

\caption{Attack success rate (ASR) of backdoor attacks before forensic analysis.}
\label{tab:asr_without_traceback}
\end{table*}

\begin{table*}[htbp]
\centering
\small
\addtolength{\tabcolsep}{-4.4pt}
\renewcommand{\arraystretch}{1.5}
\begin{tabular}{llccccccccccc}
\toprule
{Case} & {Method} & SIMPLE & COVERT & TROJAN & CB-SA & CB-GPT & CB-CGPT & BadCode & CodePoi \\
\midrule
\multirow{11}{*}{{jinja2}} & All-at-Once & $\frac{15}{80000}$ & $\frac{10}{80000}$ & $\frac{14}{80000}$ & $\frac{12}{80000}$ & $\frac{10}{80000}$ & $\frac{8}{80000}$ & $\frac{27}{80000}$ & $\frac{18}{80000}$ \\
 & Step-by-Step & $\frac{105}{80000}$ & $\frac{73}{80000}$ & $\frac{25}{80000}$ & $\frac{31}{80000}$ & $\frac{96}{80000}$ & $\frac{37}{80000}$ & $\frac{97}{80000}$ & $\frac{129}{80000}$ \\
 & Binary Search & $\frac{63}{80000}$ & $\frac{54}{80000}$ & $\frac{11}{80000}$ & $\frac{17}{80000}$ & $\frac{61}{80000}$ & $\frac{50}{80000}$ & $\frac{57}{80000}$ & $\frac{60}{80000}$ \\
 & Context-Cite & $\frac{63}{80000}$ & $\frac{55}{80000}$ & $\frac{82}{80000}$ & $\frac{68}{80000}$ & $\frac{65}{80000}$ & $\frac{51}{80000}$ & $\frac{58}{80000}$ & $\frac{56}{80000}$ \\
 & Self-Citation & $\frac{60}{80000}$ & $\frac{67}{80000}$ & $\frac{85}{80000}$ & $\frac{74}{80000}$ & $\frac{61}{80000}$ & $\frac{18}{80000}$ & $\frac{62}{80000}$ & $\frac{54}{80000}$ \\
 & PoiF & $\frac{163}{80000}$ & $\frac{172}{80000}$ & $\frac{242}{80000}$ & $\frac{198}{80000}$ & $\frac{217}{80000}$ & $\frac{208}{80000}$ & $\frac{154}{80000}$ & $\frac{172}{80000}$ \\
 & RAGForensics & $\frac{145}{80000}$ & $\frac{208}{80000}$ & $\frac{218}{80000}$ & $\frac{209}{80000}$ & $\frac{283}{80000}$ & $\frac{289}{80000}$ & $\frac{175}{80000}$ & $\frac{145}{80000}$ \\
 & TracLLM & $\frac{56}{80000}$ & $\frac{54}{80000}$ & $\frac{62}{80000}$ & $\frac{47}{80000}$ & $\frac{32}{80000}$ & $\frac{19}{80000}$ & $\frac{61}{80000}$ & $\frac{68}{80000}$ \\
 & AttnTrace & $\frac{64}{80000}$ & $\frac{48}{80000}$ & $\frac{62}{80000}$ & $\frac{49}{80000}$ & $\frac{39}{80000}$ & $\frac{17}{80000}$ & $\frac{48}{80000}$ & $\frac{51}{80000}$ \\
 & RAGOrigin & $\frac{105}{80000}$ & $\frac{187}{80000}$ & $\frac{166}{80000}$ & $\frac{144}{80000}$ & $\frac{122}{80000}$ & $\frac{94}{80000}$ & $\frac{133}{80000}$ & $\frac{111}{80000}$ \\
 \rowcolor{customred}
\cellcolor{white}
 & \alg & $\frac{40}{80000}$ & $\frac{26}{80000}$ & $\frac{52}{80000}$ & $\frac{32}{80000}$ & $\frac{36}{80000}$ & $\frac{47}{80000}$ & $\frac{51}{80000}$ & $\frac{43}{80000}$ \\
\midrule
\multirow{11}{*}[-0.3em]{{requests}} & All-at-Once & $\frac{13}{80000}$ & $\frac{9}{80000}$ & $\frac{10}{80000}$ & $\frac{18}{80000}$ & $\frac{29}{80000}$ & $\frac{15}{80000}$ & $\frac{17}{80000}$ & $\frac{15}{80000}$ \\
 & Step-by-Step & $\frac{76}{80000}$ & $\frac{20}{80000}$ & $\frac{17}{80000}$ & $\frac{28}{80000}$ & $\frac{60}{80000}$ & $\frac{35}{80000}$ & $\frac{82}{80000}$ & $\frac{73}{80000}$ \\
 & Binary Search & $\frac{20}{80000}$ & $\frac{8}{80000}$ & $\frac{5}{80000}$ & $\frac{9}{80000}$ & $\frac{24}{80000}$ & $\frac{27}{80000}$ & $\frac{18}{80000}$ & $\frac{25}{80000}$ \\
 & Context-Cite & $\frac{66}{80000}$ & $\frac{83}{80000}$ & $\frac{85}{80000}$ & $\frac{71}{80000}$ & $\frac{68}{80000}$ & $\frac{37}{80000}$ & $\frac{70}{80000}$ & $\frac{72}{80000}$ \\
 & Self-Citation & $\frac{61}{80000}$ & $\frac{72}{80000}$ & $\frac{88}{80000}$ & $\frac{76}{80000}$ & $\frac{59}{80000}$ & $\frac{48}{80000}$ & $\frac{61}{80000}$ & $\frac{68}{80000}$ \\
 & PoiF & $\frac{109}{80000}$ & $\frac{197}{80000}$ & $\frac{142}{80000}$ & $\frac{122}{80000}$ & $\frac{88}{80000}$ & $\frac{68}{80000}$ & $\frac{113}{80000}$ & $\frac{112}{80000}$ \\
 & RAGForensics & $\frac{107}{80000}$ & $\frac{142}{80000}$ & $\frac{109}{80000}$ & $\frac{98}{80000}$ & $\frac{114}{80000}$ & $\frac{223}{80000}$ & $\frac{129}{80000}$ & $\frac{107}{80000}$ \\
 & TracLLM & $\frac{59}{80000}$ & $\frac{88}{80000}$ & $\frac{80}{80000}$ & $\frac{73}{80000}$ & $\frac{55}{80000}$ & $\frac{24}{80000}$ & $\frac{52}{80000}$ & $\frac{62}{80000}$ \\
 & AttnTrace & $\frac{68}{80000}$ & $\frac{83}{80000}$ & $\frac{84}{80000}$ & $\frac{77}{80000}$ & $\frac{50}{80000}$ & $\frac{28}{80000}$ & $\frac{73}{80000}$ & $\frac{71}{80000}$ \\
 & RAGOrigin & $\frac{85}{80000}$ & $\frac{107}{80000}$ & $\frac{66}{80000}$ & $\frac{99}{80000}$ & $\frac{76}{80000}$ & $\frac{105}{80000}$ & $\frac{92}{80000}$ & $\frac{88}{80000}$ \\
  \rowcolor{customred}
\cellcolor{white}
 & \alg & $\frac{26}{80000}$ & $\frac{33}{80000}$ & $\frac{41}{80000}$ & $\frac{60}{80000}$ & $\frac{72}{80000}$ & $\frac{35}{80000}$ & $\frac{42}{80000}$ & $\frac{25}{80000}$ \\
\midrule
\multirow{11}{*}[-0.3em]{{socket}} & All-at-Once & $\frac{9}{80000}$ & $\frac{5}{80000}$ & $\frac{3}{80000}$ & $\frac{13}{80000}$ & $\frac{25}{80000}$ & $\frac{19}{80000}$ & $\frac{10}{80000}$ & $\frac{16}{80000}$ \\
 & Step-by-Step & $\frac{30}{80000}$ & $\frac{65}{80000}$ & $\frac{28}{80000}$ & $\frac{36}{80000}$ & $\frac{52}{80000}$ & $\frac{60}{80000}$ & $\frac{38}{80000}$ & $\frac{33}{80000}$ \\
 & Binary Search & $\frac{18}{80000}$ & $\frac{11}{80000}$ & $\frac{15}{80000}$ & $\frac{7}{80000}$ & $\frac{34}{80000}$ & $\frac{30}{80000}$ & $\frac{18}{80000}$ & $\frac{19}{80000}$ \\
 & Context-Cite & $\frac{74}{80000}$ & $\frac{72}{80000}$ & $\frac{87}{80000}$ & $\frac{71}{80000}$ & $\frac{65}{80000}$ & $\frac{16}{80000}$ & $\frac{71}{80000}$ & $\frac{77}{80000}$ \\
 & Self-Citation & $\frac{73}{80000}$ & $\frac{75}{80000}$ & $\frac{84}{80000}$ & $\frac{62}{80000}$ & $\frac{46}{80000}$ & $\frac{27}{80000}$ & $\frac{75}{80000}$ & $\frac{70}{80000}$ \\
 & PoiF & $\frac{276}{80000}$ & $\frac{237}{80000}$ & $\frac{277}{80000}$ & $\frac{284}{80000}$ & $\frac{305}{80000}$ & $\frac{330}{80000}$ & $\frac{270}{80000}$ & $\frac{237}{80000}$ \\
 & RAGForensics & $\frac{204}{80000}$ & $\frac{182}{80000}$ & $\frac{106}{80000}$ & $\frac{137}{80000}$ & $\frac{199}{80000}$ & $\frac{315}{80000}$ & $\frac{193}{80000}$ & $\frac{137}{80000}$ \\
 & TracLLM & $\frac{75}{80000}$ & $\frac{86}{80000}$ & $\frac{80}{80000}$ & $\frac{68}{80000}$ & $\frac{22}{80000}$ & $\frac{18}{80000}$ & $\frac{62}{80000}$ & $\frac{71}{80000}$ \\
 & AttnTrace & $\frac{69}{80000}$ & $\frac{76}{80000}$ & $\frac{83}{80000}$ & $\frac{67}{80000}$ & $\frac{29}{80000}$ & $\frac{20}{80000}$ & $\frac{65}{80000}$ & $\frac{65}{80000}$ \\
 & RAGOrigin & $\frac{96}{80000}$ & $\frac{68}{80000}$ & $\frac{94}{80000}$ & $\frac{107}{80000}$ & $\frac{82}{80000}$ & $\frac{143}{80000}$ & $\frac{98}{80000}$ & $\frac{63}{80000}$ \\
  \rowcolor{customred}
\cellcolor{white}
 & \alg & $\frac{32}{80000}$ & $\frac{60}{80000}$ & $\frac{28}{80000}$ & $\frac{46}{80000}$ & $\frac{69}{80000}$ & $\frac{52}{80000}$ & $\frac{46}{80000}$ & $\frac{52}{80000}$ \\
\bottomrule
\end{tabular}
\renewcommand{\arraystretch}{1.0}
\caption{False positive rate (FPR) of \alg and baselines under different backdoor attacks across three vulnerability cases. We use 80,000 clean fine-tuning examples in total.}
\label{FPR_all}
\end{table*}

\begin{table*}[htbp]
\centering
\addtolength{\tabcolsep}{-4.5pt}
\small
\renewcommand{\arraystretch}{1.5}
\begin{tabular}{llccccccccccc}
\toprule
{Case} & {Method} & SIMPLE & COVERT & TROJAN & CB-SA & CB-GPT & CB-CGPT & BadCode & CodePoi \\
\midrule
\multirow{11}{*}{jinja2} & All-at-Once & $\frac{79988}{80020}$ & $\frac{79991}{80020}$ & $\frac{79986}{80140}$ & $\frac{79989}{80020}$ & $\frac{79991}{80020}$ & $\frac{79992}{80020}$ & $\frac{79975}{80020}$ & $\frac{79984}{80020}$ \\
 & Step-by-Step & $\frac{79898}{80020}$ & $\frac{79930}{80020}$ & $\frac{79975}{80140}$ & $\frac{79974}{80020}$ & $\frac{79909}{80020}$ & $\frac{79975}{80020}$ & $\frac{79907}{80020}$ & $\frac{79875}{80020}$ \\
 & Binary Search & $\frac{79939}{80020}$ & $\frac{79948}{80020}$ & $\frac{79989}{80140}$ & $\frac{79988}{80020}$ & $\frac{79942}{80020}$ & $\frac{79970}{80020}$ & $\frac{79946}{80020}$ & $\frac{79942}{80020}$ \\
 & Context-Cite & $\frac{79941}{80020}$ & $\frac{79949}{80020}$ & $\frac{79920}{80140}$ & $\frac{79935}{80020}$ & $\frac{79939}{80020}$ & $\frac{79930}{80020}$ & $\frac{79946}{80020}$ & $\frac{79948}{80020}$ \\
 & Self-Citation & $\frac{79944}{80020}$ & $\frac{79936}{80020}$ & $\frac{79917}{80140}$ & $\frac{79929}{80020}$ & $\frac{79943}{80020}$ & $\frac{79964}{80020}$ & $\frac{79942}{80020}$ & $\frac{79951}{80020}$ \\
 & PoiF & $\frac{79844}{80020}$ & $\frac{79836}{80020}$ & $\frac{79808}{80140}$ & $\frac{79814}{80020}$ & $\frac{79800}{80020}$ & $\frac{79807}{80020}$ & $\frac{79854}{80020}$ & $\frac{79836}{80020}$ \\
 & RAGForensics & $\frac{79859}{80020}$ & $\frac{79800}{80020}$ & $\frac{79824}{80140}$ & $\frac{79801}{80020}$ & $\frac{79726}{80020}$ & $\frac{79762}{80020}$ & $\frac{79831}{80020}$ & $\frac{79859}{80020}$ \\
 & TracLLM & $\frac{79940}{80020}$ & $\frac{79957}{80020}$ & $\frac{79942}{80140}$ & $\frac{79956}{80020}$ & $\frac{79967}{80020}$ & $\frac{79966}{80020}$ & $\frac{79943}{80020}$ & $\frac{79935}{80020}$ \\
 & AttnTrace & $\frac{79948}{80020}$ & $\frac{79951}{80020}$ & $\frac{79942}{80140}$ & $\frac{79958}{80020}$ & $\frac{79975}{80020}$ & $\frac{79962}{80020}$ & $\frac{79957}{80020}$ & $\frac{79954}{80020}$ \\
 & RAGOrigin & $\frac{79900}{80020}$ & $\frac{79823}{80020}$ & $\frac{79861}{80140}$ & $\frac{79864}{80020}$ & $\frac{79889}{80020}$ & $\frac{79909}{80020}$ & $\frac{79872}{80020}$ & $\frac{79894}{80020}$ \\
   \rowcolor{customred}
\cellcolor{white}
 & {\alg} & $\frac{79980}{80020}$ & $\frac{79994}{80020}$ & $\frac{80085}{80140}$ & $\frac{79988}{80020}$ & $\frac{79984}{80020}$ & $\frac{79953}{80020}$ & $\frac{79969}{80020}$ & $\frac{79977}{80020}$ \\
\midrule
\multirow{11}{*}[-0.3em]{requests} & All-at-Once & $\frac{79988}{80020}$ & $\frac{79992}{80020}$ & $\frac{79990}{80140}$ & $\frac{79982}{80020}$ & $\frac{79983}{80020}$ & $\frac{79989}{80020}$ & $\frac{79984}{80020}$ & $\frac{79986}{80020}$ \\
 & Step-by-Step & $\frac{79952}{80020}$ & $\frac{79991}{80020}$ & $\frac{79983}{80140}$ & $\frac{79969}{80020}$ & $\frac{79964}{80020}$ & $\frac{79968}{80020}$ & $\frac{79922}{80020}$ & $\frac{79931}{80020}$ \\
 & Binary Search & $\frac{79982}{80020}$ & $\frac{80002}{80020}$ & $\frac{79995}{80140}$ & $\frac{79991}{80020}$ & $\frac{79992}{80020}$ & $\frac{79984}{80020}$ & $\frac{79984}{80020}$ & $\frac{79977}{80020}$ \\
 & Context-Cite & $\frac{79944}{80020}$ & $\frac{79934}{80020}$ & $\frac{79930}{80140}$ & $\frac{79938}{80020}$ & $\frac{79944}{80020}$ & $\frac{79926}{80020}$ & $\frac{79933}{80020}$ & $\frac{79931}{80020}$ \\
 & Self-Citation & $\frac{79940}{80020}$ & $\frac{79928}{80020}$ & $\frac{79929}{80140}$ & $\frac{79928}{80020}$ & $\frac{79939}{80020}$ & $\frac{79932}{80020}$ & $\frac{79943}{80020}$ & $\frac{79935}{80020}$ \\
 & PoiF & $\frac{79892}{80020}$ & $\frac{79869}{80020}$ & $\frac{79875}{80140}$ & $\frac{79898}{80020}$ & $\frac{79905}{80020}$ & $\frac{79912}{80020}$ & $\frac{79892}{80020}$ & $\frac{79894}{80020}$ \\
 & RAGForensics & $\frac{79895}{80020}$ & $\frac{79888}{80020}$ & $\frac{79891}{80140}$ & $\frac{79902}{80020}$ & $\frac{79897}{80020}$ & $\frac{79795}{80020}$ & $\frac{79875}{80020}$ & $\frac{79898}{80020}$ \\
 & TracLLM & $\frac{79941}{80020}$ & $\frac{79924}{80020}$ & $\frac{79944}{80140}$ & $\frac{79933}{80020}$ & $\frac{79945}{80020}$ & $\frac{79952}{80020}$ & $\frac{79953}{80020}$ & $\frac{79942}{80020}$ \\
 & AttnTrace & $\frac{79952}{80020}$ & $\frac{79934}{80020}$ & $\frac{79940}{80140}$ & $\frac{79937}{80020}$ & $\frac{79950}{80020}$ & $\frac{79956}{80020}$ & $\frac{79930}{80020}$ & $\frac{79932}{80020}$ \\
 & RAGOrigin & $\frac{79915}{80020}$ & $\frac{79899}{80020}$ & $\frac{79854}{80140}$ & $\frac{79905}{80020}$ & $\frac{79915}{80020}$ & $\frac{79915}{80020}$ & $\frac{79911}{80020}$ & $\frac{79914}{80020}$ \\
   \rowcolor{customred}
\cellcolor{white}
 & {\alg} & $\frac{79994}{80020}$ & $\frac{79987}{80020}$ & $\frac{80028}{80140}$ & $\frac{79960}{80020}$ & $\frac{79928}{80020}$ & $\frac{79983}{80020}$ & $\frac{79978}{80020}$ & $\frac{79995}{80020}$ \\
\midrule
\multirow{11}{*}[-0.3em]{socket} & All-at-Once & $\frac{79991}{80020}$ & $\frac{79995}{80020}$ & $\frac{79997}{80140}$ & $\frac{79987}{80020}$ & $\frac{79986}{80020}$ & $\frac{79993}{80020}$ & $\frac{79991}{80020}$ & $\frac{79985}{80020}$ \\
 & Step-by-Step & $\frac{79967}{80020}$ & $\frac{79982}{80020}$ & $\frac{79972}{80140}$ & $\frac{79966}{80020}$ & $\frac{79964}{80020}$ & $\frac{79970}{80020}$ & $\frac{79965}{80020}$ & $\frac{79970}{80020}$ \\
 & Binary Search & $\frac{79983}{80020}$ & $\frac{79993}{80020}$ & $\frac{79985}{80140}$ & $\frac{79994}{80020}$ & $\frac{79979}{80020}$ & $\frac{79980}{80020}$ & $\frac{79984}{80020}$ & $\frac{79982}{80020}$ \\
 & Context-Cite & $\frac{79936}{80020}$ & $\frac{79936}{80020}$ & $\frac{79926}{80140}$ & $\frac{79939}{80020}$ & $\frac{79940}{80020}$ & $\frac{79968}{80020}$ & $\frac{79932}{80020}$ & $\frac{79925}{80020}$ \\
 & Self-Citation & $\frac{79937}{80020}$ & $\frac{79935}{80020}$ & $\frac{79924}{80140}$ & $\frac{79940}{80020}$ & $\frac{79949}{80020}$ & $\frac{79946}{80020}$ & $\frac{79927}{80020}$ & $\frac{79933}{80020}$ \\
 & PoiF & $\frac{79758}{80020}$ & $\frac{79778}{80020}$ & $\frac{79742}{80140}$ & $\frac{79773}{80020}$ & $\frac{79724}{80020}$ & $\frac{79651}{80020}$ & $\frac{79736}{80020}$ & $\frac{79769}{80020}$ \\
 & RAGForensics & $\frac{79786}{80020}$ & $\frac{79793}{80020}$ & $\frac{79810}{80140}$ & $\frac{79809}{80020}$ & $\frac{79793}{80020}$ & $\frac{79606}{80020}$ & $\frac{79815}{80020}$ & $\frac{79869}{80020}$ \\
 & TracLLM & $\frac{79935}{80020}$ & $\frac{79925}{80020}$ & $\frac{79931}{80140}$ & $\frac{79937}{80020}$ & $\frac{79956}{80020}$ & $\frac{79962}{80020}$ & $\frac{79942}{80020}$ & $\frac{79932}{80020}$ \\
 & AttnTrace & $\frac{79941}{80020}$ & $\frac{79938}{80020}$ & $\frac{79934}{80140}$ & $\frac{79936}{80020}$ & $\frac{79942}{80020}$ & $\frac{79960}{80020}$ & $\frac{79939}{80020}$ & $\frac{79939}{80020}$ \\
 & RAGOrigin & $\frac{79904}{80020}$ & $\frac{79906}{80020}$ & $\frac{79813}{80140}$ & $\frac{79893}{80020}$ & $\frac{79929}{80020}$ & $\frac{79869}{80020}$ & $\frac{79906}{80020}$ & $\frac{79941}{80020}$ \\
   \rowcolor{customred}
\cellcolor{white}
 & {\alg} & $\frac{79988}{80020}$ & $\frac{79960}{80020}$ & $\frac{79972}{80140}$ & $\frac{79974}{80020}$ & $\frac{79951}{80020}$ & $\frac{79968}{80020}$ & $\frac{79974}{80020}$ & $\frac{79968}{80020}$ \\
\bottomrule
\end{tabular}
\renewcommand{\arraystretch}{1.0}
\caption{Detection accuracy (DACC) of \alg and baseline methods under different backdoor attacks across three vulnerability cases. Note that for the TROJAN attack, as described in~\cite{yan2024llm}, each malicious sample is expanded into seven replicated variants, producing 140 backdoored examples and a total of 80,140 fine-tuning samples, including 80,000 clean ones. For the other attacks, the attacker generates 20 backdoored examples, yielding a total of 80,020 fine-tuning samples.}
\label{DACC_all}
\end{table*}

\begin{table*}[htbp]
\centering
\addtolength{\tabcolsep}{-4.4pt}
\small 
\begin{tabular}{llcccccccc}
\toprule
{Case} & {Method} & {SIMPLE} & {COVERT} & {TROJAN} & {CB-SA} & {CB-GPT} & {CB-CGPT} & {BadCode} & {CodePoi} \\
\midrule

\multirow{11}{*}[-0.3em]{{jinja2}} 
 & All-at-Once   & 0.53 & 0.37 & 0.21 & 0.69 & 0.67 & 0.57 & 0.51 & 0.33 \\
 & Step-by-Step  & 0.50 & 0.19 & 0.18 & 0.52 & 0.52 & 0.39 & 0.47 & 0.39 \\
 & Binary Search & 0.48 & 0.31 & 0.15 & 0.53 & 0.60 & 0.50 & 0.55 & 0.33 \\
 & Context-Cite  & 0.52 & 0.17 & 0.15 & 0.57 & 0.68 & 0.54 & 0.51 & 0.40 \\
 & Self-Citation & 0.47 & 0.21 & 0.21 & 0.63 & 0.60 & 0.52 & 0.55 & 0.40 \\
 & PoiF          & 0.40 & 0.15 & 0.07 & 0.49 & 0.20 & 0.43 & 0.50 & 0.43 \\
 & RAGForensics  & 0.53 & 0.15 & 0.08 & 0.54 & 0.40 & 0.17 & 0.53 & 0.34 \\
 & TracLLM       & 0.54 & 0.13 & 0.19 & 0.57 & 0.48 & 0.57 & 0.54 & 0.38 \\
 & AttnTrace     & 0.51 & 0.22 & 0.19 & 0.55 & 0.47 & 0.56 & 0.47 & 0.34 \\
 & RAGOrigin     & 0.50 & 0.09 & 0.11 & 0.47 & 0.37 & 0.21 & 0.51 & 0.35 \\
   \rowcolor{customred}
\cellcolor{white} 
 & {\alg} 
                   & 0.02 & 0.00 & 0.00 
                   & 0.02 & 0.01 & 0.00 
                   & 0.01 & 0.01 \\

\midrule

\multirow{11}{*}[-0.3em]{{requests}} 
 & All-at-Once   & 0.85 & 0.83 & 0.24 & 0.82 & 0.64 & 0.40 & 0.85 & 0.80 \\
 & Step-by-Step  & 0.86 & 0.81 & 0.26 & 0.70 & 0.55 & 0.35 & 0.86 & 0.73 \\
 & Binary Search & 0.88 & 0.81 & 0.22 & 0.79 & 0.65 & 0.41 & 0.85 & 0.74 \\
 & Context-Cite  & 0.78 & 0.79 & 0.25 & 0.77 & 0.62 & 0.26 & 0.81 & 0.72 \\
 & Self-Citation & 0.83 & 0.80 & 0.26 & 0.78 & 0.52 & 0.27 & 0.87 & 0.80 \\
 & PoiF          & 0.73 & 0.67 & 0.23 & 0.60 & 0.50 & 0.23 & 0.70 & 0.63 \\
 & RAGForensics  & 0.83 & 0.72 & 0.25 & 0.75 & 0.58 & 0.22 & 0.81 & 0.70 \\
 & TracLLM       & 0.80 & 0.76 & 0.27 & 0.80 & 0.39 & 0.30 & 0.84 & 0.83 \\
 & AttnTrace     & 0.79 & 0.75 & 0.25 & 0.84 & 0.58 & 0.25 & 0.83 & 0.77 \\
 & RAGOrigin     & 0.88 & 0.65 & 0.20 & 0.78 & 0.53 & 0.27 & 0.79 & 0.75 \\
   \rowcolor{customred}
\cellcolor{white} 
 & {\alg} 
                   & 0.03 & 0.01 & 0.01 
                   & 0.00 & 0.00 & 0.02 
                   & 0.02 & 0.01 \\

\midrule

\multirow{11}{*}[-0.3em]{{socket}} 
 & All-at-Once   & 0.90 & 0.64 & 0.23 & 0.86 & 0.61 & 0.63 & 0.84 & 0.85 \\
 & Step-by-Step  & 0.90 & 0.64 & 0.22 & 0.78 & 0.53 & 0.48 & 0.77 & 0.77 \\
 & Binary Search & 0.90 & 0.66 & 0.23 & 0.81 & 0.62 & 0.60 & 0.80 & 0.76 \\
 & Context-Cite  & 0.84 & 0.62 & 0.18 & 0.77 & 0.58 & 0.40 & 0.81 & 0.74 \\
 & Self-Citation & 0.78 & 0.63 & 0.27 & 0.81 & 0.55 & 0.42 & 0.83 & 0.72 \\
 & PoiF          & 0.82 & 0.54 & 0.26 & 0.71 & 0.51 & 0.47 & 0.84 & 0.55 \\
 & RAGForensics  & 0.80 & 0.58 & 0.23 & 0.73 & 0.56 & 0.32 & 0.70 & 0.74 \\
 & TracLLM       & 0.85 & 0.66 & 0.22 & 0.79 & 0.57 & 0.42 & 0.85 & 0.75 \\
 & AttnTrace     & 0.93 & 0.64 & 0.19 & 0.74 & 0.60 & 0.38 & 0.83 & 0.78 \\
 & RAGOrigin     & 0.77 & 0.49 & 0.20 & 0.75 & 0.57 & 0.34 & 0.81 & 0.83 \\
   \rowcolor{customred}
\cellcolor{white} 
 & {\alg} 
                   & 0.02 & 0.00 & 0.00 
                   & 0.02 & 0.00 & 0.00 
                   & 0.02 & 0.01 \\

\bottomrule
\end{tabular}
\caption{ASR of backdoor attacks after removing backdoored examples identified by different forensics methods.}
\label{tab:asr_after_traceback}
\end{table*}

\begin{table*}[t]
\small 
\centering
\addtolength{\tabcolsep}{-4.5pt}
\begin{tabular}{l | c c c c c c c c c c c}
\toprule
Method &
\makecell{All-at\\-Once} &
\makecell{Step-by\\-Step} &
\makecell{Binary\\Search} &
\makecell{Context\\-Cite} &
\makecell{Self-\\Citation} &
PoiF &
\makecell{RAG-\\Forensics} &
\makecell{Trac-\\LLM} &
\makecell{Attn-\\Trace} &
\makecell{RAG-\\Origin} &
\cellcolor{customred}\makecell{Code\\Tracer} \\
\midrule
Time &
97.15 & 70.14 & 39.28 & 102.45 & 129.23 & 85.60 &
49.33 & 72.55 & 64.91 & 82.33 & \cellcolor{customred}47.10 \\
\bottomrule
\end{tabular}
\caption{Running time of different methods (in seconds)}
\label{tab:running_time}
\end{table*}

\begin{table*}[t]
\small 
\centering
\addtolength{\tabcolsep}{-4.5pt}
\begin{tabular}{l | c c c c c c c c c c c}
\toprule
Method &
\makecell{All-at\\-Once} &
\makecell{Step-by\\-Step} &
\makecell{Binary\\Search} &
\makecell{Context\\-Cite} &
\makecell{Self-\\Citation} &
PoiF &
\makecell{RAG-\\Forensics} &
\makecell{Trac-\\LLM} &
\makecell{Attn-\\Trace} &
\makecell{RAG-\\Origin} &
\cellcolor{customred}\makecell{Code\\Tracer} \\
\midrule
Cost &
0.31 & 0.33 & 0.28 & 0.35 & 0.35 & 0.34 &
0.34 & 0.32 & 0.35 & 0.37 & \cellcolor{customred}0.33 \\
\bottomrule
\end{tabular}
\caption{Monetary cost of different methods (in USD)}
\label{tab:monetary_cost}
\end{table*}

\begin{table*}[t]
\centering
\small 
\begin{tabular}{l|ccccc}
\toprule
{Trigger} & {SIMPLE} & {COVERT}  & {CB-SA} & {CB-GPT} & {CB-CGPT} \\
\midrule
Random code        & 0.01 & 0.01 & 0.00 & 0.01 & 0.00 \\
Targeted code        & 0.03 & 0.01 & 0.02 & 0.00 & 0.00 \\
\bottomrule
\end{tabular}
 \caption{FNR of \alg for different triggers for jinja2.}
 \label{tab:diff_triggers}

\end{table*}

\begin{table*}[t]
\centering
\addtolength{\tabcolsep}{-2.9pt}
\begin{tabular}{l|cccccccc}
\toprule
{top-$K$} & {SIMPLE} & {COVERT} &
{TROJAN} &
{CB-SA} & {CB-GPT} & {CB-CGPT} & {BadCode} & {CodePoi}\\
\midrule
250  & 0.07 & 0.05 & 0.09 & 0.07 & 0.09 & 0.05 & 0.06 & 0.07\\
500  & 0.01 & 0.01 & 0.03 & 0.01 & 0.01 & 0.00 & 0.01 & 0.03\\
1000 & 0.00 & 0.00 & 0.02 & 0.01 & 0.01 & 0.00 & 0.01 & 0.02\\
2000 & 0.02 & 0.01 & 0.04 & 0.01 & 0.02 & 0.02 & 0.03 & 0.04\\
5000 & 0.03 & 0.01 & 0.04 & 0.01 & 0.03 & 0.02 & 0.04 & 0.04\\
\bottomrule
\end{tabular}%

 \caption{Results of \alg for the different top-$K$ in the case of jinja2.}
\label{tab:topk}
\end{table*}

\begin{table}[t]
\centering
\small 
\addtolength{\tabcolsep}{-0.9pt}
\begin{tabular}{l|cccccccc}
\toprule
{Metric} & {SIMPLE} & {COVERT} &
{TROJAN} &
{CB-SA} & {CB-GPT} & {CB-CGPT} & {BadCode} & {CodePoi}\\
\midrule
FNR  & 0.02 & 0.01 & 0.02 & 0.01 & 0.01 & 0.00 & 0.02 & 0.00\\
FPR  & 0.00 & 0.00 & 0.00 & 0.00 & 0.00 & 0.00 & 0.00 & 0.00\\
DACC & 1.00 & 1.00 & 1.00 & 1.00 & 1.00 & 1.00 & 1.00 & 1.00\\
\bottomrule
\end{tabular}%
 \caption{Results of \alg for the large-scale fine-tuning dataset in the case of jinja2.}
\label{tab:large_scale}
\end{table}

\begin{table}[t]
\centering
\small 
\addtolength{\tabcolsep}{-2pt}
\begin{tabular}{l|cccccccc}
\toprule
{Variant} & {SIMPLE} & {COVERT} & TROJAN & {CB-SA} & {CB-GPT} & {CB-CGPT} & {BadCode} & {CodePoi}\\
\midrule
{Variant I}   & 0.33 & 0.51 & 0.34 & 0.32 & 0.37 & 0.39 & 0.34 & 0.31\\
{Variant II}  & 0.56 & 0.43 & 0.95 & 0.40 & 0.54 & 0.46 & 0.48 & 0.40\\
{Variant III} & 0.50 & 0.40 & 0.85 & 0.57 & 0.42 & 0.46 & 0.40 & 0.49\\
\rowcolor{customred}\alg & 0.01 & 0.01 & 0.03 & 0.01 & 0.01 & 0.00 & 0.01 & 0.03\\
\bottomrule
\end{tabular}
 \caption{FNR for variants of \alg in the case of jinja2.}
\label{tab:variants} 
\end{table}

\begin{table}[h!]
\centering
\small 
\addtolength{\tabcolsep}{-4.5pt}

\begin{tabular}{l|c c c c c c c c}
\toprule
{LLM Models} & SIMPLE & COVERT & TROJAN & CB-SA & CB-GPT & CB-CGPT & BadCode & CodePoi\\
\midrule
GPT-4o & 0.03 & 0.01 & 0.01 & 0.01 & 0.01 & 0.02 & 0.02 & 0.00\\
GPT-4.1-mini & 0.00 & 0.01 & 0.02 & 0.02 & 0.02 & 0.00 & 0.01 & 0.01\\
GPT-5 & 0.01 & 0.01 & 0.00 & 0.00 & 0.00 & 0.01 & 0.01 & 0.00\\
GPT-5-mini & 0.03 & 0.03 & 0.02 & 0.01 & 0.00 & 0.00 & 0.02 & 0.03\\
Llama-3.1-70B-Instruct & 0.03 & 0.00 & 0.03 & 0.01 & 0.01 & 0.02 & 0.04 & 0.02 \\
\bottomrule
\end{tabular}

\caption{FNR of \alg using different external LLMs in the case of jinja2.}
\label{tab:llms}  
\end{table}

\begin{table}[t]
\small 
\setlength{\tabcolsep}{3pt} 
\resizebox{\columnwidth}{!}{
\begin{tabular}{l|cccccccc}
\toprule
\text{Method} & {SIMPLE} & {COVERT} & TROJAN & {CB-SA} & {CB-GPT} & {CB-CGPT} & {BadCode} & {CodePoi}\\
\midrule
CodeQL               & 0.53 & 1.00 & 1.00 & 0.83 & 1.00 & 1.00 & 0.47 & 0.56\\
LLM detection        & 0.77 & 0.83 & 0.89 & 0.85 & 0.67 & 0.64 & 0.68 & 0.62\\
K-means  & 0.73 & 0.67 & 0.91 & 0.83 & 0.77 & 0.82 & 0.69 & 0.67\\
Spectral & 0.91 & 0.83 & 0.72 & 0.88 & 0.87 & 0.55 & 0.88 & 0.85\\
\bottomrule
\end{tabular}
}
\caption{FNR of detection methods in the case of jinja2.}
\label{tab:defense_comparison} 
\end{table}

\begin{table}[t]
\small 
\addtolength{\tabcolsep}{-3.8pt} 

\begin{tabular}{l|cccccccc}
\toprule
{Method} & SIMPLE & COVERT & TROJAN & CB-SA & CB-GPT & CB-CGPT & BadCode & CodePoi\\
\midrule
OpenRCA               & 0.63 & 0.85 & 0.78 & 0.53 & 0.75 & 0.76 & 0.59 & 0.57\\
LOCALIZEAGENT         & 0.45 & 0.61 & 0.52 & 0.58 & 0.66 & 0.71 & 0.41 & 0.43\\
\bottomrule
\end{tabular}

\caption{FNR of different software failure localization methods in the case of jinja2.}
\label{tab:existing_traceback}
\end{table}

\begin{table}[t]
\centering
\small 
\addtolength{\tabcolsep}{-3.8pt}
\begin{tabular}{lccccccccccc}
\toprule
{Attack} & \makecell{{All-at}\\{-Once}} & \makecell{{Step-by}\\{-Step}} & \makecell{{Binary}\\{Search}} & \makecell{{Context}\\{-Cite}} & \makecell{{Self-}\\{Citation}} & {PoiF} & \makecell{{RAG-}\\{Forensics}} & \makecell{{Trac-}\\{LLM}} & \makecell{{Attn-}\\{Trace}} & \makecell{{RAG-}\\{Origin}} & \cellcolor{customred} {\alg} \\
\midrule
EP & 0.95 & 0.83 & 0.91 & 0.82 & 0.77 & 0.46 & 0.51 & 0.78 & 0.68 & 0.79 & \cellcolor{customred}0.05 \\
AP & 0.94 & 0.67 & 0.83 & 0.78 & 0.79 & 0.39 & 0.47 & 0.64 & 0.67 & 0.59 & \cellcolor{customred}0.02 \\
\bottomrule
\end{tabular}

\caption{FNR of different methods under adaptive attacks for jinja2.} 
\label{tab:adaptive}
\end{table}

\begin{table}[t]
\centering
\small 

\begin{tabular}{l|c c c}
\toprule
\small 
{Metric} & SIMPLE+COVERT+TROJAN & CB-SA+CB-GPT+CB-CGPT & {BadCode+CodePoi} \\
\midrule
FNR & 0.02 & 0.01 & 0.01\\
FPR & 0.00 & 0.00 & 0.00\\
DACC & 1.00 & 1.00 & 1.00\\
\bottomrule
\end{tabular}

 \caption{Results of \alg with multi-attacker scenarios in the case of jinja2.}
\label{tab:multipleattacks}
\end{table}

\begin{table}[t]
\centering
\small 

\begin{tabular}{l|c c c c}
\toprule
{Metric} & CB-CGPT+Others & CB-GPT+Others & CB-SA+Others & COVERT+Others \\
\midrule
FNR  & 0.02 & 0.01 & 0.02 & 0.01 \\
FPR  & 0.00 & 0.00 & 0.00 & 0.00 \\
DACC & 1.00 & 1.00 & 1.00& 1.00 \\
\bottomrule
\end{tabular}

 \caption{Results of \alg with unrelated backdoored examples in the case of jinja2.}
\label{tab:unrelated_poisoned}
\end{table}

\begin{table}[t]
\centering
\small
\setlength{\tabcolsep}{4pt} 
\begin{tabular}{l|c c c c c c c c}
\toprule
{Metric} & SIMPLE & COVERT & TROJAN & CB-SA & CB-GPT & CB-CGPT & BadCode & CodePoi \\
\midrule
FNR & 0.01 & 0.01 & 0.02 & 0.01 & 0.01 & 0.00 & 0.01 & 0.00\\
FPR & 0.00 & 0.00 & 0.00 & 0.00 & 0.00 & 0.00 & 0.00 & 0.00\\
DACC & 1.00 & 1.00 & 1.00 & 1.00 & 1.00 & 1.00 & 1.00 & 1.00\\
\bottomrule
\end{tabular}

\caption{Results of \alg when the malicious completion comprises the safe completion in the case of jinja2.}
\label{tab:safe_unsafe}
\end{table}

\begin{figure*}[htbp]
  \centering
  \includegraphics[width=1.0\textwidth]{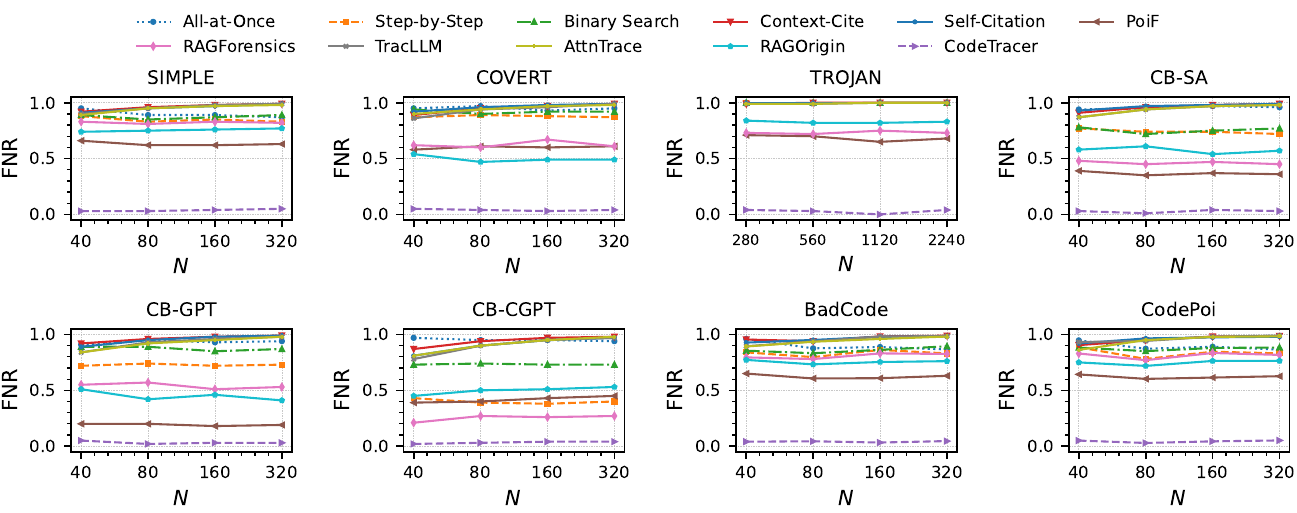}
  \caption{Impact of different numbers of backdoored examples $N$ in the case of jinja2.}
  \label{fig:acc_asr_plots}
\end{figure*}

\begin{table}[htbp]
\centering
\small 
\addtolength{\tabcolsep}{-3.5pt}
\begin{tabular}{lccccccccccc}
\toprule
{Metric} & \makecell{{All-at}\\{-Once}} & \makecell{{Step-by}\\{-Step}} & \makecell{{Binary}\\{Search}} & \makecell{{Context}\\{-Cite}} & \makecell{{Self-}\\{Citation}} & {PoiF} & \makecell{{RAG-}\\{Forensics}} & \makecell{{Trac-}\\{LLM}} & \makecell{{Attn-}\\{Trace}} & \makecell{{RAG-}\\{Origin}} & \cellcolor{customred}{\alg} \\
\midrule
FNR & 0.81 & 0.79 & 0.84 & 0.64 & 0.70 & 0.41 & 0.67 & 0.64 & 0.75 & 0.58 & \cellcolor{customred}0.02 \\
\bottomrule
\end{tabular}
\caption{FNR of \alg under the multi-hop attack in the case of jinja2.}
\label{tab:multihop}  
\end{table}

\begin{table}[t]
\centering
\small 
\setlength{\tabcolsep}{4pt} 

\begin{tabular}{l|c c c c c c c c}
\toprule
{Metric} & SIMPLE & COVERT & TROJAN & CB-SA & CB-GPT & CB-CGPT & BadCode & CodePoi\\
\midrule
FNR & 0.03 & 0.01 & 0.03 & 0.01 & 0.01 & 0.00 & 0.03 & 0.02\\
FPR & 0.00 & 0.00 & 0.00 & 0.00 & 0.00 & 0.00 & 0.00 & 0.00\\
DACC & 1.00 & 1.00 & 1.00 & 1.00 & 1.00 & 1.00 & 1.00 & 1.00\\
\bottomrule
\end{tabular}

\caption{Results of \alg with noisy user-reported malicious completion in the case of jinja2.}
\label{tab:noisy}
\end{table}

\begin{table}[t]
\centering
\small 

\begin{tabular}{l|c c}
\toprule
{Metric} & All-at-Once & Binary Search \\
\midrule
Agent-level accuracy & 53.97 & 44.83\\
\bottomrule
\end{tabular}

  \caption{Agent-level accuracies of All-at-Once and Binary Search on the Who\&When dataset~\cite{zhang2025agent}.}
\label{result_reproduce}  
\end{table}